\documentclass[twocolumn,superscriptaddress,preprintnumbers,amsmath,amssymb,pre]{revtex4}
\usepackage{graphicx}
\usepackage{dcolumn,xcolor}
\usepackage[normalem]{ulem}
\usepackage{amsmath} 
\usepackage{amssymb}
\usepackage{amsfonts}
\usepackage{bm}
\usepackage[latin1]{inputenc}
\usepackage[dvips]{epsfig}
\usepackage{hyperref}
\usepackage{comment}

\usepackage{color}
\usepackage{tikz}
\usetikzlibrary{shapes}

\newcommand{\circlelightgoldenrod}{\raisebox{0.5pt}{\tikz{\node[draw,scale=0.5,circle,fill=white!60!yellow](){};}}}

\newcommand{\circledarkgoldenrod}{\raisebox{0.5pt}{\tikz{\node[draw,scale=0.5,circle,fill=black!0!brown,rotate=0](){};}}}

\newcommand{\circlesiennafour}{\raisebox{0.5pt}{\tikz{\node[draw,scale=0.5,circle,fill=black!40!brown](){};}}}

\newcommand{\triangledarkgoldenrod}{\raisebox{0.5pt}{\tikz{\node[draw,scale=0.3,regular polygon, regular polygon sides=3,fill=black!0!brown,rotate=0](){};}}}

\newcommand{\diamondsiennafour}{\raisebox{0.5pt}{\tikz{\node[draw,scale=0.4,diamond,fill=black!40!brown](){};}}}

\newcommand{\squarebrownfour}{\raisebox{0.5pt}{\tikz{\node[draw,scale=0.4,regular polygon, regular polygon sides=4,fill=black!60!red](){};}}}

\newcommand{\squarefigeleven}{\raisebox{0.5pt}{\tikz{\node[draw,scale=0.4,regular polygon, regular polygon sides=4,fill=black!70!brown](){};}}}

\newcommand{\violetcircle}{\raisebox{0.5pt}{\tikz{\node[draw,scale=0.5,circle,fill=black!20!violet](){};}}}

\newcommand{\grayscurocircle}{\raisebox{0.5pt}{\tikz{\node[draw,scale=0.5,circle,fill=black!20!gray](){};}}}

\newcommand{\circlefigeleven}{\raisebox{0.5pt}{\tikz{\node[draw,scale=0.5,circle,fill=black!20!brown](){};}}}

\newcommand{\redinvertedtriangle}{\raisebox{0pt}{\tikz{\node[draw,scale=0.4,regular polygon, regular polygon sides=3,fill=black!10!red,rotate=180](){};}}}

\newcommand{\invertedtrianglefigeleven}{\raisebox{0pt}{\tikz{\node[draw,scale=0.4,regular polygon, regular polygon sides=3,fill=orange!50!pink,rotate=180](){};}}}

\newcommand{\greentriangle}{\raisebox{0.5pt}{\tikz{\node[draw,scale=0.3,regular polygon, regular polygon sides=3,fill=black!10!green,rotate=0](){};}}}

\newcommand{\graytriangle}{\raisebox{0.5pt}{\tikz{\node[draw,scale=0.3,regular polygon, regular polygon sides=3,fill=black!80!white,rotate=0](){};}}}

\newcommand{\graysquare}{\raisebox{0.5pt}{\tikz{\node[draw,scale=0.55,regular polygon, regular polygon sides=4,fill=white!70!gray](){};}}}

\newcommand{\cyansquare}{\raisebox{0.5pt}{\tikz{\node[draw,scale=0.4,regular polygon, regular polygon sides=4,fill=black!10!cyan](){};}}}

\newcommand{\yellowdiamond}{\raisebox{0pt}{\tikz{\node[draw,scale=0.45,diamond,fill=black!10!yellow](){};}}}

\newcommand{\orangediamond}{\raisebox{0pt}{\tikz{\node[draw,scale=0.45,diamond,fill=black!0!orange](){};}}}

\newcommand{\reddiamond}{\raisebox{0pt}{\tikz{\node[draw,scale=0.4,diamond,fill=black!10!red](){};}}}

\definecolor{fig1_1}{rgb}{1.000000000000000 ,  0.781200000000000  , 0.497500000000000}
\definecolor{green(ncs)}{rgb}{0.0, 0.62, 0.42}
\definecolor{deepskyblue}{rgb}{0.0, 0.75, 1.0}

\newcommand{\dd}{\mathrm{d}}
\newcommand{\gp}{\dot\gamma}
\newcommand{\gm}{\gamma_\text{M}}

\newcommand{\sm}{\sigma_{\text{M}}}

\newcommand{\sy}{\sigma_\text{y}}
\newcommand{\Gp}{\dot\Gamma}

\newcommand{\tit}{\tilde{t}}
\newcommand{\tm}{t_\text{M}}
\newcommand{\titm}{\tit_\text{M}}
\newcommand{\titone}{\tit_\text{1}}

\newcommand{\Sm}{\Sigma_\text{M}}

\newcommand{\Tf}{T_\text{f}}

\begin{document}

\title{Continuum modelling of shear start-up in soft glassy materials}

  \author{Roberto Benzi}
 \affiliation{Dipartimento di Fisica, Universit\`a di Roma ``Tor Vergata" and INFN, Via della Ricerca Scientifica, 1-00133 Roma, Italy\looseness=-1}
 \author{Thibaut Divoux}
\affiliation{Univ Lyon, Ens de Lyon, Univ Claude Bernard, CNRS, Laboratoire de Physique, F-69342 Lyon, France\looseness=-1}
\author{Catherine Barentin}
  \affiliation{Universit\'e de Lyon, Universit\'e Claude Bernard Lyon 1, CNRS, Institut Lumi\`ere Mati\`ere, F-69622 Villeurbanne, France\looseness=-1}
 \author{S\'ebastien Manneville}
\affiliation{Univ Lyon, Ens de Lyon, Univ Claude Bernard, CNRS, Laboratoire de Physique, F-69342 Lyon, France\looseness=-1}
\author{Mauro Sbragaglia}
\affiliation{Dipartimento di Fisica, Universit\`a di Roma ``Tor Vergata" and INFN, Via della Ricerca Scientifica, 1-00133 Roma, Italy\looseness=-1}
\author{Federico Toschi}
\affiliation{Department of Applied Physics, Eindhoven University of Technology, P.O. Box 513, 5600 MB Eindhoven, The Netherlands and CNR-IAC, Rome, Italy}

\date{\today}

\begin{abstract}
Yield stress fluids (YSFs) display a dual nature highlighted by the existence of a critical stress $\sy$ such that YSFs are solid for stresses $\sigma$ imposed below $\sy$, whereas they flow like liquids for $\sigma > \sy$. Under an applied shear rate $\gp$, the solid-to-liquid transition is associated with a complex spatiotemporal scenario that depends on the microscopic details of the system, on the boundary conditions, and on the system size. Still, the general phenomenology reported in the literature boils down to a simple sequence that can be divided into a short-time response characterized by the so-called ``stress overshoot", followed by stress relaxation towards a steady state. Such relaxation can be either {($i$)} long-lasting, which usually involves the growth of a shear band that can be only transient or that may persist at steady-state, or {($ii$)} abrupt, in which case the solid-to-liquid transition resembles the failure of a brittle material, involving avalanches. In the present paper, we use a continuum model based on a spatially-resolved fluidity approach to rationalize the complete scenario associated with the shear-induced yielding of YSFs. A key feature of our model is to provide a scaling for the coordinates of the stress overshoot, i.e., stress $\sm$ and strain $\gm$ as a function of $\gp$, which shows good agreement with experimental and numerical data extracted from the literature. Moreover, our approach shows that the power-law scaling $\sm(\gp)$ is intimately linked to the growth dynamics of a fluidized boundary layer in the vicinity of the moving boundary. Yet, such scaling is independent of the fate of that layer, and of the long-term behavior of the YSF, i.e., whether the steady-state flow profile is homogeneous or shear-banded. Finally, when including the presence of ``long-range" correlations, we show that our model displays a ductile to brittle transition, i.e., the stress overshoot reduces into a sharp stress drop associated with avalanches, which impacts the scaling $\sm(\gp)$. This generalized model nicely captures subtle avalanche-like features of the transient shear banding dynamics reported in experiments. Our work offers a unified picture of shear-induced yielding in YSFs, whose complex spatiotemporal dynamics are deeply connected to non-local effects.  
\end{abstract}

\maketitle

\section{Introduction}

Yield stress fluids (YSFs) encompass a wide variety of amorphous soft materials, from soft glasses like shaving creams and mayonnaise to colloidal gels such as fresh cement pastes and silica or alumina precursors for catalyst supports. These materials all share the existence of a critical stress $\sy$, coined the yield stress, below which their mechanical response is mainly that of a solid, and above which they flow like liquids~\cite{Barnes:1999,Bonn:2017,Coussot:2018}. Such shear-induced solid-to-liquid transition is ``reversible" in the sense that the material's solid-like behavior is recovered when shear is stopped --although generally with a different microstructure from that preceding the transition~\cite{Koumakis:2015,Helal:2016,Narayanan:2017,Jamali:2020}. As a consequence, the mechanical properties of YSFs at a given point in time are shaped by their shear history, which makes modelling particularly challenging while also yielding fruitful developments in the field~\cite{Balmforth:2014,Fielding:2014,Nicolas:2018,Wei:2018,Larson:2019}. 

Due to their intrinsic viscoelastic nature, YSFs are characterized by complex yielding dynamics in both time and space. For instance, in a shear start-up condition, i.e., when imposing a constant shear rate $\gp$ from rest at time $t=0$, YSFs display a non-monotonic stress response. At short times, the system first shows a linear elastic stress response to the applied strain $\gamma(t) = \int_0^t \gp(s) \,{\rm d}s=\gp t$, i.e., the shear stress $\sigma(t)$ grows proportionally to $\gamma(t)$. The stress then reaches a maximum $\sm$ at time $\tm$, corresponding to a strain $\gm=\gp\tm$, and then further decreases towards its steady-state value. This time sequence is usually referred to as the ``stress overshoot" in the literature~\cite{Mewis:2009}. 
At $\gamma = \gm$, the stress maximum coincides with a pronounced anisotropy of the YSF microstructure~\cite{Mohraz:2005}, while for $t>\tm$, the stress relaxation coincides with the fluidization of the material, which translates at the microscale into local plastic events, such as cage breaking and super-diffusive particle motion in glasses~\cite{Zausch:2008,Koumakis:2012,Laurati:2017}, and failure of strands in gels~\cite{Masschaele:2009,Rajaram:2010,Rajaram:2011}. 
Moreover, at small scales, the stress relaxation comes with a broad variety of flow profiles, from a ductile-like response involving shear banding, either transient or permanent~\cite{Mas:1994,Ovarlez:2009,Divoux:2010,Divoux:2012}, to a more abrupt and brittle-like rupture that can either take place in the bulk or at the boundary of the shearing device~\cite{Magnin:1990,Persello:1994,Pignon:1996,Divoux:2011}. The subsequent long-term, steady-state response corresponds to the flow curve $\sigma_{\rm ss}(\gp)$, which is usually well-described by the empirical Herschel-Bulkley (HB) law:
\begin{equation}\label{eq:HB}
\sigma_{\rm ss} (\dot \gamma) = \sy + A \dot \gamma ^ n\,,
\end{equation}
where $A$ and $n$ respectively denote the consistency index and a shear-thinning exponent generally close to $0.5$~\cite{Barnes:2001,Cloitre:2003,Becu:2006,Katgert:2009,Dinkgreve:2015}.

From a modeling perspective, various approaches have been explored to capture the distinctive features of the yielding transition, i.e., the stress overshoot and the subsequent occurrence of shear banding, either only transient or persisting at steady state. The stress overshoot has been observed in Molecular Dynamics (MD) and Brownian Dynamics (BD) simulations, which have confirmed the growth of microstructural anisotropy under shear up to the stress maximum, and provided insights on the local scenario of the yielding transition beyond the stress maximum for both gels and glasses~\cite{Whittle:1997,Varnik:2004,Park:2013,Santos:2013,Colombo:2014,Park:2017,Johnson:2018}. Moreover, Mode Coupling Theory has shown the robustness of the stress overshoot, which may arise even in the presence of homogeneous flows in well-equilibrated, i.e., non-aging systems~\cite{Zausch:2008,Amann:2013,Amann:2014,Amann:2015}. Finally, the stress overshoot was also captured by 2D mesoscopic elastoplastic models \cite{Jagla:2007}, the Soft Glassy Rheology (SGR) model~\cite{Moorcroft:2011}, the Shear Transformation Zone (STZ) model~\cite{Manning:2007} and fluidity models~\cite{Derec:2003,Moorcroft:2011,deSouzaMendes:2013,Fielding:2014}. This latter class of models is also known as ``structural kinetic theories'', and often referred to as ``$\lambda$-models'', named after the structural parameter $\lambda$ of the fluid used to characterize the state of its microstructure. Since their introduction in the late 1930s, $\lambda$-models have been refined to encompass thixotropic elasto-viscoplastic effects, among which aging, and also include advanced concepts from the theory of plasticity such as kinematic hardening to capture shear-induced memory effects in YSFs in their solid state~\cite{Dimitriou:2013,Dimitriou:2014,Geri:2017,Wei:2019,Larson:2019}.

Except for a few contributions based on MD simulations of binary Lennard-Jones glasses that report a logarithmic scaling for the amplitude of the stress overshoot $\sm$ with the applied shear rate $\gp$~\cite{Varnik:2004,Rottler:2005,Shrivastav:2016}, most of the models listed above conclude that $\sm$ scales as a power law of $\gp$, with an exponent 0.5. However, confronting theory, simulations and experiments is made difficult due to the lack of consensus on the relevant quantity that allows one to compare different types of YSFs, e.g., gels and glasses with different yield stresses. For instance, experimental and numerical literature indiscriminately report $\sm$~\cite{Whittle:1997,Derec:2003,Carrier:2009,Divoux:2011,Koumakis:2011,Park:2013}, $\sm-\sigma_{0}$ where $\sigma_{0}$ is a reference stress chosen arbitrarily beyond the overshoot~\cite{Dimitriou:2014}, $\sm/\sy-1$~\cite{Letwimolnun:2007}, or $\sm/\sigma_{\rm ss}-1$~\cite{Koumakis:2012,Amann:2013,Amann:2014,Koumakis:2016,Sentjabrskaja:2018} as a function of $\gp$. 
As for the flow profile, transient shear bands have been reported recently in both fluidity and SGR models~\cite{Moorcroft:2011}, STZ models~\cite{Hinkle:2016}, athermal local yield stress models derived from MCT~\cite{Liu:2018b}, and MD simulations, in which long-lasting transient heterogeneous flows were linked to the presence of over-constrained microscopic domains~\cite{Vasisht:2020,Vasisht:2020b,Golkia:2020}. Steady-state shear banding has been extensively described by the models discussed above. For the liquid-to-solid transition, steady-state shear banding is observed below a critical shear rate that results from the competition between spontaneous aging of the YSF microstructure and rejuvenation by shear~\cite{Ovarlez:2009,Divoux:2016}. However, in the case of the solid-to-liquid transition, such as in shear start-up experiments, the steady-state flow profile strongly depends on the sample age, the nature of the boundary conditions and the geometry~\cite{Gibaud:2008,Seth:2012,Vayssade:2014}. These effects have been discussed in the framework of tensorial elasto-visco-plastic approaches~\cite{Cheddadi:2012}. 

Given the complexity of the aforementioned features, it is a challenging task to build a theoretical framework capable of providing a unified view of the yield stress transition. Only a few theoretical approaches propose a consistent description of the yielding transition that encompasses both the stress overshoot and the subsequent homogeneous or heterogeneous flow profiles~\cite{Fielding:2014}. Therefore, a unified description of the yielding transition still stands out as an open challenge, including questions such as the following ones: (i) what sets the amplitude of the stress overshoot? (ii) is there any link between the stress overshoot and the subsequent flow dynamics, in particular the existence of a (transient) shear band? (iii) how does the presence of spatial correlations in the flow affect the stress overshoot and the long-term dynamics of YSFs?

In previous works, some of us have revisited a spatially-resolved fluidity model, devised originally in Ref.~\cite{Bocquet:2009}, that takes into account ``non-local effects'' to describe the steady-state flow behavior of YSFs~\cite{Benzi:2016,Lulli:2018}. Our model is built on the assumption that the steady-state flow of the YSF is described by the HB law [see Eq.~\eqref{eq:HB}] and this is approached via a dynamical equation for the local fluidity $f$. The fluidity is introduced as the basic coarse-grained degree of freedom for a YSF and is related to the plastic activity in the system, i.e., it is non-zero only when non-negligible plastic events occurs. 
As discussed in Ref.~\cite{Bocquet:2009}, the spatial heterogeneity of $f$ is controlled by a ``cooperative scale'' $\xi$, which relates to the extension of the region that is impacted by a neighboring plastic rearrangement: 
other plastic events may occur nearby with $\xi$ representing the corresponding correlation length, which is typically of the order of a few times the size of the elementary constituents of the YSF microstructure~\cite{Goyon:2008,Geraud:2013,Geraud:2017,Nicolas:2018}. This feature is referred to as ``non-local effects'' in the literature \cite{Mansard:2012}. 
Our model assumes a dynamical equation for $f$ where two rheological branches exist: a solid branch corresponding to $f=0$ and a fluid branch corresponding to $f>0$. 

Recently, we have made two significant steps forward in modelling the complete yielding scenario based on the above fluidity model. First, we have shown that our theoretical approach captures long-lasting transient shear-banded flows for YSFs that exhibit homogeneous flow in steady state~\cite{Benzi:2019}. For these YSFs, otherwise known as ``simple'' YSFs~\cite{Ovarlez:2013b}, our approach yielded a quantitative prediction for the power-law scaling of the strain-induced fluidization time, $\Tf\sim \gp^{-9/4}$, and provided a deep connection between the HB phenomenological exponent $n$ and the exponent for stress-induced fluidization at constant stress $\sigma$, $\Tf\sim (\sigma/\sy-1)^{-9/4n}$. Second, in the companion paper~\cite{companion2}, we have used our model to obtain scaling laws for the stress overshoot with $\gp$. More specifically, we have identified that the relevant quantity for characterizing the overshoot amplitude is $\sm/\sy-1$, which shows two asymptotic scalings, $(\sm/\sy-1)\sim\gp^{2n/3}$ at ``small" $\gp$ and $(\sm/\sy-1)\sim\gp^{4n/(9-n)}$ at ``large" $\gp$, here again providing a connection between the steady-state rheology and the transient response.
These predictions are intimately linked to the growth dynamics of a fluidized boundary layer in the vicinity of the moving boundary.
 In that framework, the spatial coexistence of a fluidized band and a solid region is the consequence of the first-order transition occurring at the shear band interface, where $\nabla f $ shows a rather sharp change.

In the present manuscript, we first generalize our model to investigate the impact of  permanent shear bands, both on the early-time response and on the overshoot scenario described in the companion paper~\cite{companion2}. Theoretical arguments supported by numerical computations show that the scaling of the stress maximum $\sm(\dot \gamma)$ is robust and independent of the fate of the fluidized layer. In a second step, we include long-range correlations, i.e., correlations over length scales larger than the cooperative scale $\xi$, and examine their influence on the yielding scenario. 
We find that the presence of long-range correlations is responsible for an increased amount of local plasticity, yielding an abrupt, avalanche-like shear-induced fluidization of the material. The presence of avalanches also leads to an earlier fluidization of the sample. Yet, the fluidization occurs over a timescale that obeys the same power-law scaling as the one measured in the absence of long-range correlations. In other words, the presence of long-range correlations only affects the prefactor of the fluidization law, which appears as characteristic of the material. Furthermore, we demonstrate that, within our generalized approach, the nature of the boundary condition for the fluidity at the moving wall strongly affects the scaling of the stress maximum associated with the stress overshoot. In particular, imposing a zero fluidity gradient at the moving wall, which prevents the growth of any fluidized band at the wall, triggers a fluidization in the bulk that involves avalanches. In practice, the stress overshoot shows an abrupt stress drop reminiscent of a brittle failure, while its maximum displays a logarithmic increase with the applied shear rate that strongly contrasts with the power-law scaling observed in the absence of long-range correlations. 

The paper is organized as follows. In Section~\ref{Theory_Homogeneous}, we recall our theoretical framework, namely a non-local fluidity approach that predicts the scalings of both the overshoot coordinates and the fluidization time with respect to the applied shear rate, for ``simple" YSFs that display a homogeneous flow profile in steady state. We then present and discuss the corresponding numerical results in details. In Section~\ref{Theory_shearbanding}, we discuss the scalings obtained with a non-local fluidity model with noise, which captures the case of YSFs that display steady-state shear banding, as well as the corresponding numerical results. In Section~\ref{sec:literatureresults}, we illustrate the robustness of the theoretical predictions obtained in the first two sections, as well as their limits, by revisiting experimental and numerical data from the literature. Section~\ref{sec:generalization} presents a  generalization of our model that includes the effects of long-range correlations and ``avalanches'', leading to a thorough understanding of the various experimental features of shear start-up in YSFs. Finally, Section~\ref{sec:discussion_conclusion} gathers the discussion and conclusion.

\section{Theoretical framework for systems with homogeneous steady state}
\label{Theory_Homogeneous}

In the following, we first recall the physical ingredients for our theoretical description of ``simple'' YSFs that show a transient shear-banding regime prior to homogeneous shear flow at steady state. We then introduce the full set of equations and derive scaling arguments for the overshoot at short times. Finally, we describe the long-time evolution towards steady state. In all cases, the dynamical equations are solved numerically to support the scaling arguments and to provide a detailed view of the evolution of the local flow field.

\subsection{Non-local fluidity model for transient shear banding}

\subsubsection{Physical ingredients and assumptions underlying the model}

Physically, we aim at modelling the situation where a constant shear rate $\dot\gamma$ is applied to a YSF confined between two parallel walls separated by a distance $L$. Such a one-dimensional shear flow is characterized by a velocity field $\mathbf{v}(x,y,t)=v(y,t)\mathbf{e}_x$, where $x$ is the velocity direction and $y$ is the velocity gradient direction. It is obtained by imposing a constant velocity $v_0$ to the moving boundary located at $y=0$, i.e., $v(0,t)=v_0$, while the boundary at $y=L$ is held fixed, i.e., $v(L,t)=0$ at all times $t$. As recalled in the introduction, the fluidity $f(y,t)$ relates to the  local plastic activity in the system and its dynamics involve the cooperativity length $\xi$ as a key physical parameter. In steady state, the fluidity coincides with the inverse of the local viscosity, i.e., $f=\dot\gamma/\sigma=(\partial v/\partial y)/\sigma$. In our model, we focus on a dynamical equation for $f(y,t)$ so that, rather than imposing the velocity at the walls, we will devise boundary conditions for $f$ at $y=0$ and $y=L$ together with an initial condition $f(y,0)$ that accounts for the initial preparation of the system, another key ingredient in modelling shear start-up of YSFs. 
Moreover, the dynamical equation for $f$ accounts for a steady-state rheology described by the HB law [see Eq.~\eqref{eq:HB}] through two rheological branches, namely a solid branch corresponding to $f=0$ and a fluid branch corresponding to $f>0$. Finally, in order to model the viscoelastic response of the material under start-up of shear, an elastic modulus $G_0$ should be considered in the whole dynamical scenario. The stress $\sigma$ is assumed to be spatially homogeneous and its dynamics to depend only on the imposed shear rate $\dot \gamma$, the elastic modulus $G_0$ and the space-averaged fluidity $\langle f \rangle=(1/L)\int_0^L f(y,t)\,\text{d}y$, which is a function of time $t$ only. 
\subsubsection{Set of coupled equations for fluidity and stress, \& boundary and initial conditions}
\label{sec:coupled_eq_BC}

We consider a one-dimensional system of size $L$, with $y \in [0,L]$ the spatial coordinate along the velocity gradient direction. We assume that the constitutive rheological law is given by the HB relation [see Eq.~\eqref{eq:HB}]. Furthermore, we work with the dimensionless stress and shear rate, respectively defined as $\Sigma=\sigma/\sy$ and $\Gp=\gp/(\sy/A)^{1/n}$, resulting in the dimensionless HB law: 
\begin{equation}\label{eq:simplifiedHB}
\Sigma(\Gp)=1+\Gp^{n}\,.
\end{equation}
As introduced in~\cite{Benzi:2016}, it is assumed that the flow properties of the system are governed by a functional $F[f]=\int_0^L \Phi(f,m,\xi)\,\dd y$ with
\begin{equation}\label{3}
\Phi(f,m,\xi)=\frac{1}{2} \xi^2 (\nabla f)^2-\frac{1}{2} m f^2+\frac{2}{5}f^{5/2}\,,
\end{equation}
where $f=f(y,t)$ is the dimensionless fluidity, $\xi$ the cooperativity length, and
\begin{equation}\label{4}
m=m(\Sigma)=\frac{(\Sigma-1)^\frac{1}{2n}}{\sqrt{\Sigma}}\,\Theta(\Sigma-1)\,,
\end{equation}
with $\Theta$ the Heaviside function. The HB law~\eqref{eq:simplifiedHB} is recovered as the minimum of the bulk contribution $-\frac{1}{2} m f^2+\frac{2}{5}f^{5/2}$ in Eq.~\eqref{3}, which corresponds to $f=m^2$.

In the case of shear-induced fluidization, i.e., when the system is driven by a constant shear rate $\Gp$, we have argued in~\cite{Benzi:2019} that one should consider the rescaled variable $\tilde f= f/\Gp$ rather than $f$. This rescaling, together with $\tilde{m}=m/\Gp^{1/2}$ and $\tilde{y}=\Gp^{1/4} y/\xi$ allows for a homogeneous rescaling of the functional $\Phi(f,m,\xi)=\Gp^{5/2}\Phi(\tilde{f},\tilde{m})$, where $\Phi(\tilde{f},\tilde{m})=\frac{1}{2} (\tilde{\nabla} \tilde{f})^2-\frac{1}{2} \tilde{m} \tilde{f}^2+\frac{2}{5}\tilde{f}^{5/2}$. Assuming that the system reaches the stable configuration corresponding to the minimum of $F[\tilde{f}]$, we introduce a mobility function $k(\tilde{f})$ that drives the dynamics of $\tilde{f}$, leading to
\begin{equation} \label{1}
\frac{\partial \tilde f}{\partial t} = -{\Gp }^ {5/2} k(\tilde f) \frac{\delta F [\tilde{f}]}{\delta \tilde{f}} \,.
\end{equation}
We now take the simplest non-trivial form for the mobility, $k(\tilde{f})=\tilde{f}$ and also use the rescaled time $\tit \equiv \gp t$. Under a constant shear rate, $\tit$  simply corresponds to the strain $\gamma(t)$ applied to our system. Coming back to $f = \tilde f \Gp$, we obtain:
\begin{equation}\label{6}
\frac{\partial f}{\partial \tit} =  f  \left[ \xi^ 2  \Delta  f + m f -  f ^ {3/2} \right ]\,,
\end{equation}
where $m(\Sigma)$ is given by Eq.~\eqref{4}. To close our set of equations for shear-induced fluidization, we need an evolution equation for the stress $\Sigma$, which is assumed to be homogeneous in space as posited before. Again for the sake of simplicity, we assume that the stress dynamics follow a Maxwell equation:
\begin{equation}
\label{7}
\frac{\dd \Sigma }{\dd \tit} = \frac{1}{\Gp \tau} \left( \Gp-  { \langle f \rangle \Sigma}\right)\,,
\end{equation}
where $\tau$ is the dimensionless relaxation time. Physically, $\tau$ is inversely proportional to the elastic modulus $G_0$ and is linked to the material stiffness such that, for a given shear rate, decreasing $\tau$ corresponds to increasing the stiffness.

Finally, having defined the equations of motion, we need to set the boundary conditions as well as the initial conditions. In all our numerical computations, we assume a spatially homogeneous initial condition $f(y,0)=f(0)$. Moreover, we deduce from Eq.~\eqref{7} that the minimum shear rate for the overshoot to occur reads $\Gp_0=f(0)$. We choose $f(0)=10^{-4}$ in all computations and thus perform shear start-up flows for $\Gp \ge 2\times 10^{-4}$ to trigger overshoots (except in Sect.~\ref{sec:overshoot_brittle} where other values of $f(0)$ will be specified). As for the boundary conditions, we assume $\partial_y f |_{y=L}=0$ at the fixed wall for any $\Sigma$. At  the moving wall, i.e., $y=0$, we assume that,  for $\Sigma >1$, the fluidity $f(0,\tit)$ is equal to some ``wall fluidity'' $f_w$. In the following, we make the simplest assumption $f_w= m^2$, which guarantees the existence of a stationary solution $f=m^2$ when the system reaches complete fluidization. We shall emphasize that the results discussed in the following sections are still valid as long as the wall fluidity is proportional to $m^2$. Finally, in Section~\ref{sec:generalization}, we discuss a different boundary condition at the moving wall, namely $\partial_yf|_{y=0}=0$. The physical meaning of the two boundary conditions is as follows. On the one hand, $\partial_y f=0$ implies that the boundary does not change the rate of plastic events with respect to the bulk of the system. On the other hand, $f=f_w$ states that the rate of plastic events is fixed by the boundary independently of the bulk dynamics. In both cases, the shear rate at the boundary is fixed by the external forcing [see Eq.~\eqref{vlocal} below]. 

\subsubsection{Scale invariance}

An interesting property of Eqs.~\eqref{6} and~\eqref{7} is the invariance under the scale transformation:
\begin{eqnarray}\label{10}
m \rightarrow \lambda m \\
\nonumber
f \rightarrow \lambda^2 f \\
\nonumber
\Sigma \rightarrow \lambda^2  \Sigma\\
\nonumber
\Gp \rightarrow \lambda^4 \Gp \\
\nonumber
\tit \rightarrow \lambda^{-3} \tit \\
\nonumber 
\tau \rightarrow \lambda^{-5}  \tau \\
\nonumber
\frac{\xi}{y} \rightarrow \lambda^{1/2} \frac{\xi}{y}
\end{eqnarray}
In the following, we will use the above {\it scale} invariance property of Eqs.~\eqref{6} and~\eqref{7} to discuss the scaling behaviors of the various observables.

\subsubsection{Numerical implementation and first results}
\label{numerical}
In the next sections, we shall validate our theoretical analysis against numerical computations. To this aim, we discretize Eq.~\eqref{6} on a regular grid of $N_y=512$ grid points choosing $L=1$ and different values of $\tau$ ranging from 0.01 to 10. A crucial requirement is that the value of $\xi$ needs to be much larger than the spacing $\delta y = 1/512$. In our case we used $\xi=0.04 \approx 20 \,\delta y$. When the condition $\xi \gg \delta y$ is fulfilled, then it is possible to use any good finite difference scheme which guarantees an accuracy up to order $(\delta y)^2$ together with an accurate time integration method. In our case, we used an Euler-Cauchy implicit method. When needed [see Sec.~\ref{Theory_shearbanding}], noise effects are added at the end of the deterministic integration using a Gaussian random variable with variance proportional to $\sqrt{\delta t}$, $\delta t$ being the time step of integration. Beside the initial conditions discussed above, we assume an HB index $n=1/2$, except for some specific cases in Sec.~\ref{sub:generalized} where we will also consider values of $n$ between 0.3 and 0.7 as observed in experiments.

\begin{figure}[t]
\centering
\includegraphics[width=1.0\columnwidth]{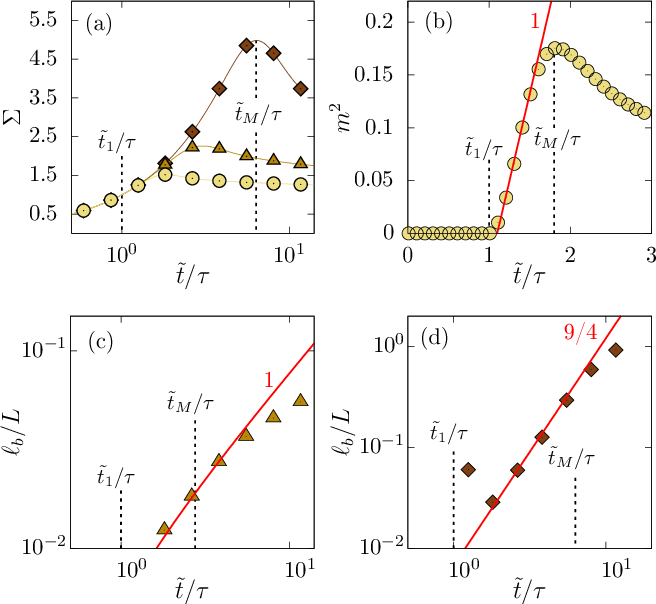}
\caption{(a) Stress response $\Sigma$ as a function of $\tit/\tau$ for $\Gp=0.004$ (\protect \circlelightgoldenrod), $0.02$ (\protect \triangledarkgoldenrod) and $5.68$~(\protect \diamondsiennafour). 
(b)~$m^2$ as a function of $\tit/\tau$ for $\Gp = 0.004$. The red line shows an affine behavior consistent with $m^2 \sim (\tit-\titone)/\tau$ (solid line) [see Eq.~\eqref{12}]. (c)~Size of the fluidized band $\ell_b$ as a function of  $\tit/\tau$ for $\Gp = 0.02$. The solid line corresponds to the affine scaling predicted in the low-shear (diffusive) regime [see Eq.~\eqref{16}]. (d) $\ell_b$ as a function of  $\tit/\tau$ for $\Gp = 5.68$. The solid line shows the power-law scaling expected in the high-shear regime with exponent $9/4$ [see Eq.~\eqref{18}]. Numerical results obtained with $n=1/2$, $\tau=10$, $\xi = 0.04$ and $L=1$.}
\label{figpreliminary0}
\end{figure}

Moreover, in order to compute velocity profiles, we need to connect the fluidity inferred from the numerical resolution of Eqs.~\eqref{6} and \eqref{7} to the local shear rate $\Gp_{\rm loc}$. The latter quantity is defined as: 
\begin{equation}
\label{vlocal}
\Gp_{\rm loc}(y,\tit) = \Gp + \left[f(y,\tit)-\langle f \rangle\right] \Sigma\,.
\end{equation}
Such a definition is justified by the fact that it leads to the following local version of Eq.~\eqref{7}:
\begin{equation}
\label{3local}
\frac{\dd\Sigma}{\dd \tit} = \frac{1}{\Gp \tau} [ \Gp_{\rm loc} - f(y,\tit)\Sigma]\,.
\end{equation}
Velocity profiles $v(y,t)$ are then computed by integration of $\Gp_{\rm loc}(y,\tit)$. Note that Eq.~\eqref{vlocal} implies that the material velocity at $y=0$ is exactly equal to that of the moving wall, $v_0=\Gp L$, at all dimensionless times $\tilde t$.

As a first example of our numerical computations, Fig.~\ref{figpreliminary0}(a) shows the stress responses $\Sigma$ versus $\tit/\tau$ for $\tau=10$ and for three different values of the shear rate $\Gp$. All responses show a clear stress overshoot that reaches larger values $\Sm$ at later times for increasing values of $\Gp$. 
We introduce $\titone$ as the dimensionless time at which $\Sigma(\titone)=1$, and $\titm$ the dimensionless time at which the stress reaches its maximum value, i.e., $\Sigma(\titm)=\Sm$ [see Fig.~\ref{figpreliminary0}(a)]. For $\tit \in [0,\titone]$ the stress remains smaller than the yield stress ($\Sigma < 1$) and $m(\Sigma) =0$. During this time interval, the fluidity at the wall $f_w=m^2$ is equal to zero [see Fig.~\ref{figpreliminary0}(b)]. Beyond $\tit=\titone$, $m^2$ increases up to a maximum reached at $\tit=\titm$, which coincides with the maximum of the stress overshoot. Finally, for $\tit \gtrsim \titm$, both the stress $\Sigma$ and the fluidity at the wall $f_w=m^2$ decrease towards their steady-state values.   
Computing the velocity profiles during shear start-up reveals that an unstable shear band develops near the moving wall, as discussed in the companion paper~\cite{companion2}. 
The size $\ell_b$ of the sheared region increases with time well beyond $\tit=\titm$  [see Fig.~\ref{figpreliminary0}(c,d)]. The shear band occupies a significant fraction of the gap width $L$, before the sample eventually experiences complete fluidization, i.e., $\ell_b=L$~\cite{Benzi:2019}. 

\subsection{Short-time local dynamics: growth of a fluidized band}\label{sub:ShortTime}
 
Let us now focus on the dynamics for $\tit \ge \titone$. At $\tit=\titone$ the value of $m$ becomes positive and a thin layer of non-zero fluidity grows near the moving wall  ($y=0$). If we assume that the term $\langle f \rangle \Sigma$ in the r.h.s.~of Eq.~\eqref{7} is small, then the stress grows linearly with time above $\titone$, i.e.,
\begin{equation}\label{11}
\Sigma(\tit)-1=\left(\frac{\tit-\titone}{\tau} \right).
\end{equation}
Obviously, this approximation does not hold for all $\tit$ above $\titone$, for which $\Sigma(\tit)$ should eventually increase in a non-linear way to satisfy $\dd \Sigma/\dd \tit=0$ at the stress maximum. 
Furthermore, for $n=1/2$ and assuming that $\Sigma$ is reasonably larger than the yield stress, i.e., $\Sigma \gg 1$, during the time interval $[\titone,\titm]$, Eq.~\eqref{4} reads $m(\Sigma)=(\Sigma-1)/\Sigma^{1/2} \sim \Sigma^{1/2} \sim (\Sigma-1)^{1/2}$. Based on Eq.~\eqref{11}, we obtain the following scaling
\begin{equation}\label{12}
m(\tit) \sim \left(\frac{\tit-\titone}{\tau}\right)^{1/2}.
\end{equation} 
Equation~\eqref{12} is numerically checked for $\Gp = 0.004$ in Fig.~\ref{figpreliminary0}(b), where  the fluidity at the wall $f_w=m^2$ indeed shows a linear growth as a function of $\tit-\titone$. At the stress maximum ($\dd \Sigma/\dd \tit =0$), the r.h.s.~of Eq.~\eqref{7} is zero, which yields
\begin{equation}\label{13}
\Gp=\left[\langle f \rangle \Sigma \right]_{\tit=\titm}\,.
\end{equation}
Moreover, the spatial average $\langle f \rangle$ is dominated by a  small fluidized region of width $\ell_b$ near the boundary, which is forced by the boundary condition $f_w = m^2$. Therefore, we may write $\langle f \rangle  \simeq m^2 \, \ell_b\,/L$, which allows us to infer $\ell_b(\tit)$ from our numerical computations. If we further estimate $m$ and $\Sigma$ using Eqs.~\eqref{11} and \eqref{12} and consistently assume $\Sm\gg 1$, such that $\Sm \sim \Sm -1$, the stress overshoot condition in Eq.~\eqref{13} becomes
\begin{equation}\label{14}
\Gp \sim \left(\Sm-1\right)^2 \, \ell_b(\titm)\,.
\end{equation}
To further extract the functional relation $\Sm(\Gp)$, we need to specify the time evolution of the band size $\ell_b$ and to evaluate it at $\tit=\titm$. We shall now discuss the growth dynamics of the shear band depending on the applied shear rate, based on the two asymptotic regimes $\Gp\ll 1$ and $\Gp\gg 1$. 

\subsubsection{Band growth controlled by diffusion}
For $\Gp\ll 1$, one may argue that in the early stages of the dynamics, the system has to ``adapt'' to the boundary condition and has to ``shape up'' the fluidized front that will later propagate through the gap. As discussed in the Appendix of Sect.~\ref{app:lsa} in full details, assuming that the diffusive term in Eq.~\eqref{6} plays a dominant role leads to the following scaling prediction for the size of the band:
\begin{equation}\label{15}
\ell_b(\tit)   \sim \sqrt{m^2(\tit) \,\xi^2\,\tit}\,.
\end{equation}
Using Eqs.~\eqref{11},~\eqref{12} and~\eqref{15}, consistently with the condition $\tit \gg \titone$, we obtain
\begin{equation}\label{16}
\ell_b(\tit) \sim \xi \tau^{1/2} \left(\frac{\tit-\titone}{\tau}\right) \sim \xi \tau^{1/2} \left(\Sigma(\tit)-1\right)\,. 
\end{equation}
Fig.~\ref{figpreliminary0}(c) shows that for $\Gp=0.02$, $\ell_b$ grows linearly with $\tit/\tau$ around the overshoot as predicted by Eq.~\eqref{16}, which supports the diffusive approximation for $\Gp\ll 1$.
\subsubsection{Band growth controlled by front propagation}
For $\Gp\gg 1$, a propagation front emerges from Eq.~\eqref{6}. Useful information on the scaling of the front speed as a function of $m$ can be inferred from the scaling properties listed in Eq.~\eqref{10}. Indeed, Eq.~\eqref{10} implies that $\dd \ell_b/\dd \tit$ must be proportional to the ratio of the characteristic scale $\xi/m^{1/2}$ to the time scale $m^{-3}$ so that
\begin{equation}\label{17}
\frac{\dd \ell_b}{\dd \tit} \sim \xi m^{5/2}\,.
\end{equation}
The validity of Eq.~\eqref{17} is discussed in more details in the Appendix of Sect.~\ref{app:lsa} based on a linear stability analysis. Using Eqs.~\eqref{12} and \eqref{17}, we obtain
\begin{equation}\label{18}
\ell_b(\tit) \sim \xi \tau  \left( \frac{\tit-\titone}{\tau} \right)^{9/4} \sim \xi \tau \left(\Sigma(\tit)-1\right)^{9/4}\,.   
\end{equation}
In Fig.~\ref{figpreliminary0}(d), we show that the size of the shear band indeed follows a power-law scaling with exponent 9/4 for $\Gp = 5.68$ around the stress overshoot.
\subsection{Scaling of the stress overshoot coordinates with applied shear rate}

We now combine the condition for the stress overshoot given in Eq.~\eqref{14} with the scalings for $\ell_b$ found for $\Gp \ll 1$ and $\Gp \gg 1$ in Eqs.~\eqref{16} and~\eqref{18} respectively. We recall that these two predictions are obtained under radically different scenarios: for $\Gp \ll 1$, the imposed shear is small and the unstable shear band does not propagate as a front but the system rather rearranges close to the boundaries, while for $\Gp \gg 1$, we used the physical ingredient of a propagating front.

\subsubsection{Diffusive regime at low shear rate}
For $\Gp \ll 1$, we combine Eq.~\eqref{14} with the scaling relation in Eq.~\eqref{16} evaluated at $\titm $, which leads to
\begin{equation}
\Gp \sim (\Sm-1)^2 \ell_b \sim \xi \tau^{1/2} (\Sm-1)^{3}\,, 
\end{equation}
so that
\begin{equation}\label{20}
\Sm-1 \sim \left( \frac{ \Gp }{ \xi \tau^{1/2} } \right)^{1/3}\,.
\end{equation}

\subsubsection{Asympototic regime at high shear rate}
For $\Gp \gg 1$, combining Eq.~\eqref{14} with the scaling relation in Eq.~\eqref{18} evaluated at $\titm$ leads to
\begin{equation}
\Gp \sim (\Sm-1)^2 \ell_b \sim \xi \tau (\Sm-1)^{17/4} \,,
\end{equation}
so that
\begin{equation}\label{22}
\Sm-1 \sim \left( \frac{ \Gp }{ \xi \tau } \right)^{4/17}\,.
\end{equation}

\subsubsection{Complete scaling for $n=1/2$}

By combining the two different scaling regimes predicted above for $\Gp \ll 1$ and $\Gp \gg 1$, we may propose the following general expression for the stress maximum $\Sm$ reached during the overshoot as a function of the applied shear rate $\Gp$:
\begin{equation}\label{23}
\Sm  -1 \sim  B \left( \frac{ \Gp }{ \xi \tau^{1/2} } \right)^{1/3} + C   \left( \frac{ \Gp }{ \xi \tau } \right)^{4/17}\,,
\end{equation}
where $B$ and $C$ are two constants independent of the model parameters. For a given $\xi$, the intersection between the two regimes occurs at a characteristic shear rate $\Gp^*$, whose scaling relation with $\tau$ can be found by balancing the two terms 
$(\Gp/\xi\tau^{1/2})^{1/3}$ and $(\Gp/\xi\tau)^{4/17}$ in Eq.~\eqref{23}. This yields
\begin{equation}\label{24}
\Gp^* \sim  \xi \tau^{-0.7}\,.
\end{equation}
When $\Gp=\Gp^*$, both terms in Eq.~\eqref{23} lead to the same scaling $\Sm^*-1 \sim \tau^{-0.4}$. Thus, we expect the overshoot data to collapse onto a master curve when plotting  $(\Sm-1)\tau^{0.4}$ as a function of $\Gp/\Gp^*$. In other words, this means that 
\begin{equation}\label{general_result}
\Sm- 1 = \tau^{-0.4} \mathcal{G}[ \Gp / \Gp^*]\,,
\end{equation}
where $\mathcal{G}(x)$ is a universal function such that $\mathcal{G}(x) \sim x^{1/3}$ for  $x\ll 1$ and $\mathcal{G}(x) \sim x^{4/17} $ for  $x\gg 1$.

\subsubsection{Generalization for any HB exponent $n$}\label{sub:generalized}

\begin{figure}[b]
\centering
\includegraphics[width=0.45\textwidth]{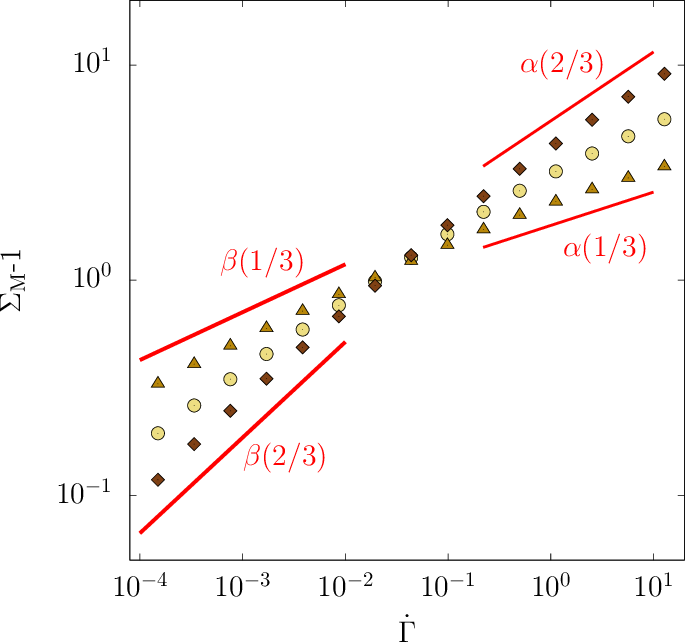}
\caption{Stress maximum $\Sm-1$ as a function of $\Gp$ for three different values of the exponent $n$ in the HB constitutive relation $\Sigma = 1+ \Gp^n$, namely $n=1/3$ (\protect \triangledarkgoldenrod) $n=1/2$ (\protect \circlelightgoldenrod) and $n=2/3$ (\protect \diamondsiennafour). The straight lines show the different scaling predictions, respectively at low shear rates with exponent $\beta(n)$ (diffusive regime) and at high shear rates with exponent $\alpha(n)$ (asymptotic regime) [see Eq.~\eqref{alphan}]. The numerical computations are performed with $\tau=10$, $\xi=0.04$ and $L=1$ for all cases.}
\label{figureovershootn}
\end{figure}

So far, we have assumed that the HB exponent $n$ takes the specific value $n=1/2$ to derive Eq.~\eqref{general_result} from Eqs.~\eqref{6} and~\eqref{7}. Here, we generalize the above results for any value of $n$. Starting from Eq.~\eqref{4}, which reads $m^2(\Sigma) = (\Sigma-1)^{1/n}/\Sigma$ for $\Sigma>1$, and following the same steps as in the previous paragraphs, it is easily shown that Eq.~\eqref{23} is changed to:
\begin{equation}
\label{result_stress_n}
\Sm -1   \sim  B \left( \frac{ \Gp }{ \xi \tau^{1/2} } \right)^{\beta(n)} + C   \left( \frac{ \Gp }{ \xi \tau } \right)^{\alpha(n)}\,,
\end{equation}
where
\begin{equation}
\label{alphan}
\beta(n) = \frac{2n}{3} \,\,\,\,\text{and}\,\,\,\, \alpha(n) = \frac{4n}{9-n}\,.
\end{equation}
Thus, the generalized version of Eq.~\eqref{general_result} takes the form:
\begin{equation}
\label{general_result_n}
\Sm -1 = \tau^{-\alpha(n)[1+\lambda(n)]}\, \mathcal{G}_n\left[ \frac{ \Gp \tau^{\lambda(n) }}{ \xi } \right]\,,
\end{equation}
where
\begin{equation}
\label{general_result_lambda}
\lambda(n)= \frac{\alpha(n)-\frac{\beta(n)}{2}}{\beta(n)-\alpha(n)}=\frac{3+n}{6-2n}\,,
\end{equation}
and $\mathcal{G}_n$ is the equivalent of $\mathcal{G}$ in Eq.~\eqref{general_result} such that $\mathcal{G}_n(x) \sim x^{\beta(n)}$ for  $x\ll 1$ (diffusive regime) and $\mathcal{G}(x) \sim x^{\alpha(n)} $ for  $x\gg 1$ (asymptotic regime). For $n=1/2$, we check that $\beta(1/2)=1/3$ and $\alpha(1/2)=4/17$, so that we recover $\lambda(1/2) = 0.7$ as in Eq.~\eqref{24}. Finally, to make the link with the additional exponent $\mu$ used in the companion paper \cite{companion2}, one has:
$\mu(n)= \alpha(n)[1+\lambda(n)]=2n/(3-n).$

In the companion paper, the validity of Eq.~\eqref{general_result} has been ascertained for $n=1/2$ and various values of $\tau$ (see Fig.~2 in Ref.~\cite{companion2}). Here, in Fig.~\ref{figureovershootn}, we compare $\Sm-1$ versus $\Gp$ computed for $n=1/3$, $n=1/2$ and $n=2/3$, and for a given $\tau=10$. As shown by the straight lines, the scalings of Eq.~\eqref{general_result_n} provide excellent predictions for the numerical results both in the diffusive regime, $\Gp\ll\Gp^*$, and in the asymptotic regime, $\Gp\gg\Gp^*$, where $\Gp^*\sim\xi\tau^{-\lambda(n)}\simeq 0.01$. 
Note that the robustness of our general approach is further tested against experimental results on Carbopol microgels with $n=0.5$--0.6 in the companion paper (see Fig.~3 in Ref.~\cite{companion2}).

\subsection{Transient dynamics up to full fluidization}

\subsubsection{Stress evolution towards equilibrium}\label{sec:flowphasediagram}

\begin{figure}[t]
\begin{center}
\includegraphics[width=1.0\columnwidth]{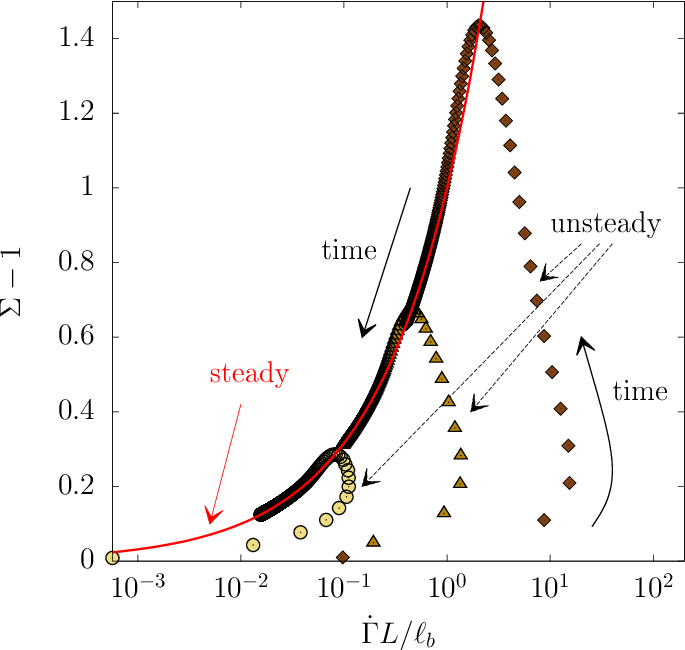}
\end{center}
\caption{Shear stress $\Sigma-1$ as a function of the effective shear rate within the shear band $ {\Gp}L/\ell_b$, where $\ell_b$ is the thickness of the  (unstable) shear band, for various global shear rates $\Gp=8\times 10^{-4}$ (\protect \circlelightgoldenrod), $\Gp=9\times 10^{-3}$ (\protect \triangledarkgoldenrod) and $\Gp=10^{-2}$ (\protect \diamondsiennafour). The solid line is the HB prediction $\Sigma =1 + (\Gp L/\ell_b)^{1/2}$ written in terms of the effective shear rate. Numerical results obtained with $n=1/2$, $\tau=10$, $\xi = 0.04$ and $L=1$. }
\label{figpreliminary1}
\end{figure}

Based on our theoretical discussion, we expect that the effective shear rate increases quite rapidly at short time scales. Once the stress maximum is reached, a slower decaying dynamics occurs with a balance between $\Sigma$, $\Gp$ and $\ell_b$ such that the r.h.s.~of Eq.~\eqref{7} is close to zero. These qualitative expectations are actually borne out by the analysis of the ``flow phase diagram'' presented in Fig.~\ref{figpreliminary1}, where we report $\Sigma(\tit)-1$ as a function of $\Gp L/\ell_b(\tit)$ during the whole fluidization process. $\Gp L/\ell_b(\tit)$ represents the effective shear rate within the shear band and the red continuous line shows the corresponding HB prediction 
\begin{equation}\label{eq:HBlocal}
\Sigma =1 + \left(\frac{\Gp L}{\ell_b}\right)^{1/2}.
\end{equation}
Such a representation of the flow dynamics highlights a clear-cut separation between two different dynamical regimes: one regime where the data fall way below the HB curve and another regime where they all collapse onto the HB curve. The former regime corresponds to the initial dynamics where $\ell_b$ is small, i.e., $\Gp L/ \ell_b$ is large, and the stress grows rapidly. This ``unsteady'' dynamical regime brings the system from the initial time up to the stress overshoot. Once the stress maximum is reached, the system enters a second dynamical regime where $\dd\Sigma/\dd \tit \sim 0$. During this ``steady'' dynamical regime, Eq.~\eqref{7} implies that $\Gp \sim \langle f \rangle \Sigma$. Using the estimation $\langle f \rangle \sim m^2 \ell_b/L \sim \Sigma \, \ell_b/L$ as in Sect.~\ref{sub:ShortTime}, we get that $(\Sigma-1)^2\sim  \Gp L/\ell_b$, which boils down the HB equation. 

\subsubsection{Early-time velocity profiles and elastic recoil around the overshoot}

\begin{figure}[t]
\begin{center}
\includegraphics[width=1.0\columnwidth]{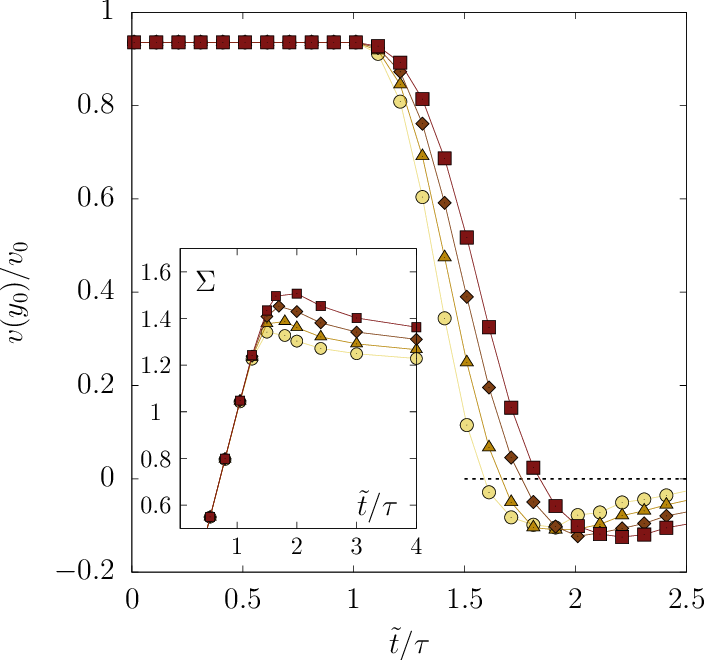}
\end{center}
\caption{Normalized velocity $v(y_0,t)/v_0$ computed near the moving wall at $y_0=L/16$ as a function of $\tit/\tau$ for $\Gp=7.61\times 10^{-4}$ (\protect \circlelightgoldenrod), $1.14 \times 10^{-3}$ (\protect \triangledarkgoldenrod), $1.71 \times 10^{-3}$ (\protect \diamondsiennafour), and $2.56\times 10^{-3}$ (\protect \squarebrownfour). $v_0$ is the velocity of the moving wall. Inset: corresponding stress response $\Sigma$ versus $\tit/\tau$. Numerical results obtained with $n=1/2$, $\tau=10$, $\xi = 0.04$ and $L=1$.}
\label{figpreliminary1bis}
\end{figure}

The temporal evolution of the normalized velocity $v(y_0,t)/v_0$ computed near the moving wall at $y_0=L/16$ is displayed in Fig.~\ref{figpreliminary1bis} for four different values of $\Gp$ and up to $\tit=2.5 \tau$, i.e., just after the stress overshoot (see inset of Fig.~\ref{figpreliminary1bis}). During, the very early stage of the stress response, the material undergoes elastic loading and its deformation remains affine. Thus, velocity profiles are linear as long as $\tit\lesssim \tau$, as shown in the companion paper \cite{companion2}. This translates in Fig.~\ref{figpreliminary1bis} into a constant velocity given by the affine displacement $v(y_0=L/16,t)=15v_0/16$ for $\tit\lesssim \tau$, independent of the shear rate $\Gp$.

Just before the overshoot, the shear band starts to nucleate and once the overshoot is reached, as $\dd\Sigma/\dd\tit <0$, the local velocity close to the moving boundary strongly decreases until a small but detectable negative velocity is measured at $y_0$ near the shear band interface. This elastic recoil is observed typically for $\tit/\tau\simeq 1.5$--3 depending on the shear rate. For larger $\tit/\tau$, the velocity at $y_0$ tends to zero, indicative of the presence of an arrested region. As also discussed in the companion paper (see Fig.~4 in Ref.~\cite{companion2}), this behavior is fully consistent with the one observed experimentally in Ref.~\cite{Divoux:2011}, where negative velocities are reported close to the moving surface together with vanishingly small velocities close to the fixed wall.

\subsubsection{Shear band evolution at long times}
\label{sec:fluidizationtime}

\begin{figure}[t]
\begin{center}
\includegraphics[width=1.0\columnwidth]{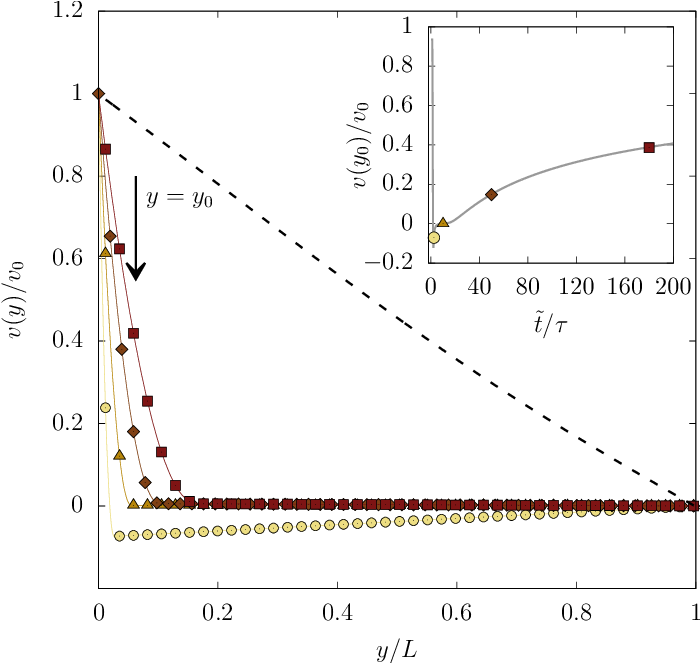}
\end{center}
\caption{Normalized velocity profile $v(y)/v_0$ for $\Gp = 1.7\times 10^{-3}$ at times $\tit/\tau=2.5$ (\protect \circlelightgoldenrod), $10$ (\protect \triangledarkgoldenrod), $50$ (\protect \diamondsiennafour) and $180$ (\protect \squarebrownfour). The dashed line shows the velocity profiles recorded at steady state. Inset: normalized velocity $v(y_0,t)/v_0$ computed near the moving wall at $y_0=L/16$ (see arrow in the main panel) as a function of $\tit/\tau$. Numerical results obtained with $n=1/2$, $\tau=10$, $\xi = 0.04$ and $L=1$.}\label{fig_preliminary2bis}
\end{figure}

We now turn to the long-term evolution of the velocity field. In Fig.~\ref{fig_preliminary2bis}, we show a selection of normalized velocity profiles $v(y)/v_0$ for $\Gp=1.7\times 10^{-3}$ at times ranging from just after the overshoot ($\tit=2.5 \tau$) up to much longer times ($\tit=180\,\tau$). After the shear band nucleation and the elastic recoil discussed above, the shear band is observed to slowly grow while the arrested region shrinks over time. This growth of the shear band leads to a slow yet steady increase of the local velocity $v(y_0=L/16)$ past the overshoot (see inset of Fig.~\ref{fig_preliminary2bis}). As shown in our previous work focused on the long-time shear-induced fluidization~\cite{Benzi:2019}, this long-lived transient shear-banded state eventually gives way to a fully fluidized state (see dashed lines in Fig.~\ref{fig_preliminary2bis}), i.e., the shear band is unstable. Such a fluidization scenario associated with a well-defined ``fluidization time'' $\Tf$ is in excellent agreement with the one observed experimentally (see also Fig.~4 in the companion paper~\cite{companion2}).

Finally, let us discuss in more detail how the scaling law for the fluidization time $\Tf$ can be obtained in the theoretical framework of Eqs.~\eqref{6} and \eqref{7}.  During the ``steady'' dynamical regime discussed in Sec.~\ref{sec:flowphasediagram}, the relation given in Eq.~\eqref{eq:HBlocal} holds. Consequently, the front velocity can be computed from Eq.~\eqref{17} with $m=(\Sigma-1)/\Sigma^{1/2}$ and $\Sigma$ given by Eq.~\eqref{eq:HBlocal}. A Taylor expansion series in $\Gp$ has a leading order in $\Gp$ that results in the following scaling law:
\begin{equation}\label{lfluid0}
\frac{\dd \ell_b}{\dd \tit} \sim \xi \left( \frac{\Gp}{\ell_b} \right)^{5/4}\,,
\end{equation}
leading to
\begin{equation}
\label{lfluid}
\ell_b( \tit) \sim  \Gp^{5/9} (\xi \tit)^{4/9}\,.
\end{equation}
Upon complete fluidization, i.e., for $\tit>\tilde{\Tf}$, $\ell_b(\tit)=L$ does not change anymore in time. We recall that $\tit$ corresponds to the strain and is therefore itself proportional to the shear rate $\Gp$. Thus, we obtain the scaling behavior of the fluidization time $\Tf$ with the shear rate $\Gp$ as
\begin{equation}
\Tf \sim \frac{1}{\xi \Gp^{9/4}}\,,
\end{equation}
which coincides with the scaling of the fluidization time discussed in Ref.~\cite{Benzi:2019}. In other words, the scaling for the fluidization time hinges on the dynamics after the overshoot maximum, where the system is approaching the asymptotic stress value and the late dynamics of the process are almost stationary.


\section{Theoretical framework for systems with permanent shear-banding}
\label{Theory_shearbanding}
In the following, we first briefly recall the physical ingredients for our theoretical description of permanent shear bands in YSFs at low enough shear rates, as well as the corresponding set of equations first introduced in Ref.~\cite{Benzi:2016}. We then derive scaling arguments for the stress maximum reached during the overshoot at short times. Finally, we describe the long-time evolution of the velocity profiles towards steady state.
\subsection{Non-local fluidity model with noise}
Our theoretical framework to account for permanent shear bands consists in adding a stochastic noise of small amplitude to the previous non-local fluidity model. Stochastic noise introduces some decorrelation between regions of different fluidity, such that an unsheared region characterized by a {\it vacuum} solution ($f_0=0$) can coexist spatially with a sheared region characterized by non-zero fluidity in steady state~\cite{Benzi:2016}. In the absence of noise, the vacuum state ($f_0=0$), i.e., the solution of Eq.~\eqref{6}, is unstable and spatial coexistence of the two solutions takes place over a finite duration. As a result, the shear band is only transient and disappears asymptotically as seen above in Sect.~\ref{sec:fluidizationtime}.
Here, stochastic noise is added to the dynamics by modifying Eq.~\eqref{6} to :
\begin{equation}\label{33}
\frac{\partial f}{\partial \tit} = f \left[ \xi^2 \Delta f + m f - f^{3/2}+\epsilon_0/4\right] + \sqrt{\epsilon_0 f} \, w(y,\tit)\,,
\end{equation}
where $w$ is a Gaussian random process $\delta$-correlated in space and time, and $\epsilon_0$ is a material parameter that allows us to tune the level of noise. Note that the limit $\epsilon_0 \rightarrow 0 $ is not singular, in the sense that we recover the results discussed in the previous section when the noise goes down to zero.

To study the properties of Eq.~\eqref{33}, it is useful to change variable and introduce the field $\phi$ defined by $f=\phi^2$. Equation~\eqref{33} then becomes:
\begin{equation}
\label{phi1}
\frac{\partial \phi}{\partial \tit}  = - \frac{\delta \mathcal{F}[\phi]}{\delta \phi} + \sqrt{\epsilon_0} w(y,\tit)\,,
\end{equation}
where the functional $\mathcal{F}[\phi]$ is given by
\begin{equation}\label{phi2}
\mathcal{F}[\phi] = 2\int_0^L \dd y \left[ \xi^2 \phi^2 (\nabla \phi)^2 -\frac{1}{4} m \phi^4 + \frac{1}{5} \phi^4 |\phi|\right]\,.
\end{equation}
For the vacuum state $\phi \simeq 0$, it was shown in Ref.~\cite{Benzi:2016} that Eq.~\eqref{phi1} reduces to:
\begin{equation}
\label{eq:renormalization}
\frac{\partial \phi}{\partial \tit} = D \Delta \phi - R \phi + m \phi^3 - \phi^3 |\phi | + \sqrt{\epsilon_0} w(y,\tit)\,,
\end{equation}
where $D$ and $R$ are {\it renormalized} coefficients induced by fluctuations that both depend on the level of noise $\epsilon_0$ and on the cooperativity length $\xi$. Due to the term $R \phi$, the noise promotes the stability of the previously unstable vacuum solution. Moreover, Eq.~\eqref{eq:renormalization} displays the characteristic features of a first-order phase transition, which occurs for $m = m(\Sigma_c) \sim R^{1/3}$, where $\Sigma_c = 1+ \Gp_c^{1/2}$. Consequently, for stresses such that $m^3 < R$, i.e., shear rates lower than the critical shear rate $\Gp_c$, a flowing region ($f \sim m^2$) and a non-flowing region ($f \simeq 0$) coexist at steady state, which corresponds to permanent shear bands~\cite{Benzi:2016}.

The condition $m^3 < R$ for the formation of a stable shear band can also be understood in terms of competing time scales. Indeed, the noise introduces a new time scale $1/R$, which sets the characteristic time for a perturbation to decay to zero. The other time scale in the model is related to $m$ and is embedded in the term $m \phi^3$ in Eq.~\eqref{eq:renormalization}. Such a term is by construction related to the bulk rheology above the yield stress, and following the scaling arguments in Eq.~\eqref{10}, the associated time scale is $m^3$. Hence, $1/m^3$ represents the characteristic time scale of the shear band instability, which can also be considered as the correlation time of the system during the fluidization phase. In this framework, the criterion $1/R<1/m^3$ for permanent shear banding implies that the ratio between the perturbation relaxation time and the fluid correlation time is smaller than 1, i.e., that any attempt of the system to flow will be damped after a relatively short time. Note that the condition of competing time scales is also a key feature of other theoretical approaches discussing the emergence of permanent shear banding~\cite{Picard:2005,Coussot:2010,Martens:2012}.

To summarize, noise leads to the existence of a critical shear rate $\Gp_c$, below which shear bands are observed in steady state and above which shear bands are only transient. The value of $\Gp_c$ depends both on the noise amplitude $\epsilon_0$ and on the correlation length $\xi$. We now turn to the short-time response of the system in the presence of such a random noise.
\subsection{Overshoot scaling}
In this section, we examine how the scaling of the stress maximum  $\Sm(\Gp)$ during the overshoot may be affected by the presence of stochastic noise and of stable shear bands. First, because of the noise, the initial condition of the space-averaged fluidity cannot be arbitrarily small. Rather, the initial fluidity is controlled by the noise, i.e., at $\tit=0$, we have $\langle f \rangle = f_0(\epsilon_0)$. As already mentioned in Section~\ref{sec:coupled_eq_BC}, this defines a characteristic shear rate $\Gp_0 \equiv f_0(\epsilon_0)$ below which there cannot be any overshoot. Second, above the critical shear rate $\Gp_c$,  i.e., when shear bands are only transient like in the absence of noise, we expect that the theoretical arguments developed in Sect.~\ref{Theory_Homogeneous} remain valid. It is therefore interesting and worthwhile to investigate the effects of the overshoot dynamics in presence of permanent shear bands, i.e., for $\Gp_0 \ll \Gp \le \Gp_c$.

To proceed, let us assume that $\Gp_0 \ll \Gp \le \Gp_c$ and that the system shows a stress overshoot of amplitude $\Sm$ at $\tit=\titm$. Since for $\Gp \le \Gp_c$, the equilibrium value of $\Sigma$ is $\Sigma_c$~\cite{Benzi:2016}, it follows that $\Sm > \Sigma_c$. The latter inequality also implies that the size of the permanent shear band $\ell_b^s (\Gp)$ must be larger than that of the shear band $\ell_b$ at  $\tit=\titm$. Over the  interval $[\tilde{t}_1,{\titm}]$, the shear band that nucleates at the wall is driven by the forcing acting at the boundary, and one expects that the growing rate of $\ell_b$ is close to the ``deterministic" behaviour previously discussed in Eqs.~\eqref{15} and \eqref{18}. Once the stress reaches its maximum value $\Sm$, the system tends to balance $\Gp$ with $\langle f \rangle \Sigma $ as in Eq.~\eqref{13}. After the overshoot, the velocity of the propagating front remains controlled by Eq.~\eqref{17} as long as $\Sigma(\tit) \gg \Sigma_c$. However, it must slow down when $\Sigma(\tit)$ becomes comparable to $\Sigma_c$ and the shear band approaches its equilibrium value $\ell_b^s(\Gp)$. As discussed below in Sect.~\ref{sec:noise_steady_state}, during this late stage, the detailed dynamics depend on the model parameters $ \xi, \epsilon_0$, and $\Gp$.

\begin{figure}[t!]
\begin{center}
\includegraphics[width=1.0\columnwidth]{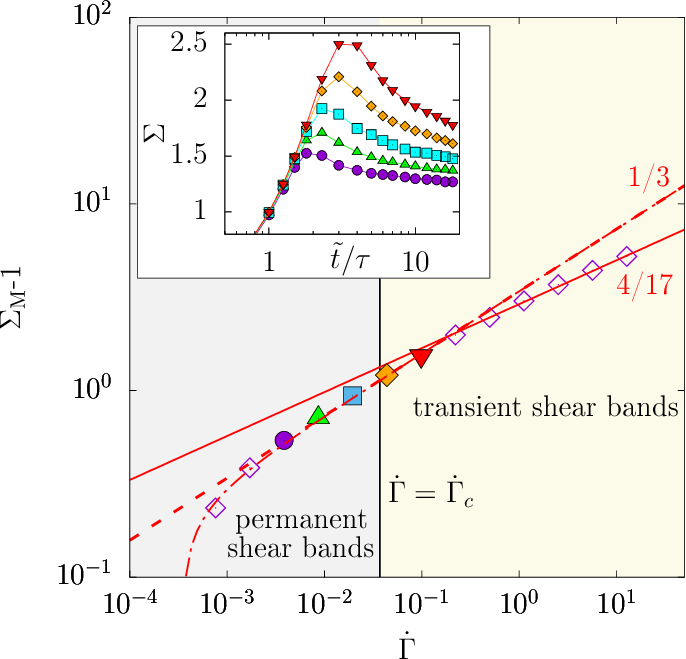}
\end{center}
\caption{
\label{figureovershootshearbanding}
Stress maximum $\Sm-1$ as a function of the applied shear rate $\Gp$ in the presence of noise. Permanent shear bands are observed for $\Gp<\Gp_c=4\times 10^{-2}$, while shear bands are only transient for $\Gp>\Gp_c$. The red dashed line shows the diffusive scaling, $\Sm-1\sim \Gp^{1/3}$, while the red solid line shows the asymptotic scaling, $\Sm-1\sim \Gp^{4/17}$. 
The dash-dotted line corresponds to Eq.~\eqref{correstions} where $\Gp_0=3.4 \times 10^{-4}$ is inferred numerically from the initial fluidity through $\Gp_0=f_0(\epsilon_0)=\langle f\rangle(\tit=0)$. Numerical results obtained by solving Eq.~\eqref{33} for $\epsilon_0= 10^{-7}$, $\tau=10$, and $\xi=0.04$. 
Inset: stress $\Sigma$ as a function of $t/\tau$ for five different shear rates: $\Gp=$0.00385 (\protect \violetcircle), 0.00865 (\protect \greentriangle), 0.0195 (\protect \cyansquare), 0.0438 (\protect \orangediamond), and 0.0985 (\protect \redinvertedtriangle) highlighted by the same symbols in the main panel. }
\end{figure}

To sum up, the scaling laws describing the locus of the stress maximum during the overshoot are not affected by the physical ingredients controlling the existence of stable or transient shear bands. In order to confirm this picture, we integrate the equation for the fluidity including noise, Eq.~\eqref{33}, for $\Gp$ in the range $[10^{-3},10]$. As in Sect.~\ref{Theory_Homogeneous}, we couple Eq.~\eqref{33} with the Maxwell-like evolution of the stress $\Sigma (\tit)$ [see Eq.~\eqref{7}], and solve both equations.
The characteristic time $\tau$ is set to 10, and the noise amplitude $\epsilon_0$ is fixed to $10^{-7}$. According to Ref.~\cite{Benzi:2016}, for $\epsilon_0=10^{-7}$ a stable shear band is observed for $3.4\times 10^{-4}\ll \Gp \le \Gp_c = 4\times 10^{-2}$. The corresponding evolution of the stress is reported as an inset in Fig.~\ref{figureovershootshearbanding} for five shear rates on both sides of the critical shear rate $\Gp_c$. A clear overshoot is observed in all cases, whether shear bands are permanent or only transient. Moreover, we report in Fig.~\ref{figureovershootshearbanding} the stress maximum  $\Sm-1$ as a function of the shear rate $\Gp$. At high shear rates, the stress maximum is well described by the scaling of the asymptotic regime, $\Sm-1\sim \Gp^{4/17}$, whereas at low shear rates, the stress maximum follows the scaling of the diffusive regime, $\Sm-1\sim \Gp^{1/3}$, as derived in the absence of noise [see Eq.~\eqref{23} and Fig.~\ref{figureovershootn}]. Therefore, numerical computations in the presence of noise confirm that the scaling of the stress maximum is not sensitive to the type of shear banding, be it permanent or transient. Finally, in the limit of extremely low values of $\Gp$, we note that the stress maximum lies slightly below the one predicted by Eq.~\eqref{general_result} in the diffusive regime (see two leftmost points in Fig.~\ref{figureovershootshearbanding}).
Such a deviation is due to the noise, which generates a non-zero fluidity $f_0(\epsilon_0)$ even for $\Gp=0$. 
In particular, for very low $\Gp\gtrsim \Gp_0$, the stress overshoot condition of Eq.~\eqref{13} should be corrected by a term $f_0(\epsilon_0)\Sm$. Equation~\eqref{14} then becomes:
\begin{equation}\label{37}
\Gp\sim f_0(\epsilon_0)\Sm + \ell_b(\Sm-1)^2 \,.
\end{equation}
Using the diffusive scaling in Eq.~\eqref{16}, 
we rewrite
\begin{equation}\label{38}
\Gp- f_0(\epsilon_0)\sim (\Sm-1) f_0(\epsilon_0) +  \left( \Sm-1\right)^3 \xi \tau^{1/2}\,.
\end{equation}
For small $f_0(\epsilon_0)$ and small $\Gp$, solving Eq.~\eqref{38} up to first order in $\Gp_0=f_0(\epsilon_0)$ leads to
\begin{equation}
\label{correstions}
\Sm-1 = \left( \frac{\Gp-\Gp_0}{\xi \tau^{1/2}}\right)^{1/3}\,.
\end{equation}
Thus, Eq.~\eqref{correstions} predicts a deviation from the diffusive regime, which is all the more pronounced that the shear rate is close to $\Gp_0$. As shown by the dash-dotted line in Fig.~\ref{figureovershootshearbanding}, Eq.~\eqref{correstions} accounts very well for $\Sm-1$ at the lowest shear rates with no fitting parameter since $\Gp_0$ is directly extracted numerically from the initial fluidity. 
\subsection{Asymptotic evolution towards steady state}
\label{sec:noise_steady_state}

\begin{figure}[!t]
\begin{center}
\includegraphics[width=1.0\columnwidth]{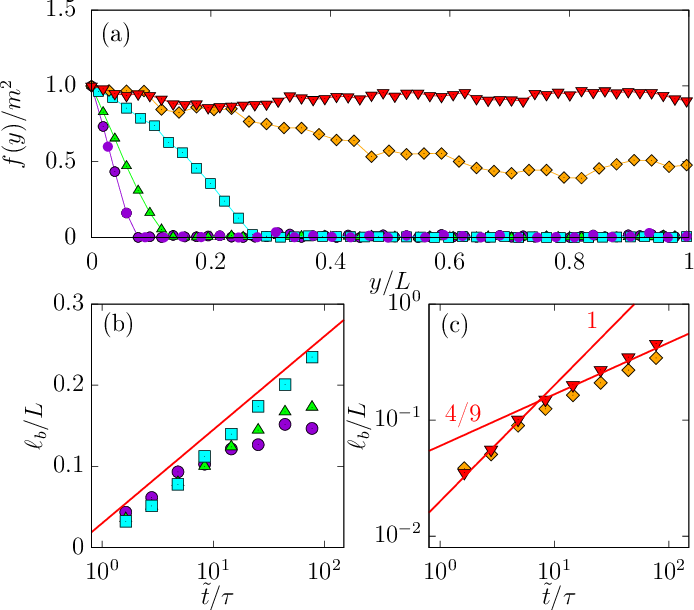}
\caption{(a) Steady-state normalized fluidity $f(y)/m^2$ as a function of $y/L$ achieved for the five shear rates reported in the inset of Fig.~\ref{figureovershootshearbanding}: $\Gp=$0.00385 (\protect \violetcircle), 0.00865 (\protect \greentriangle), 0.0195 (\protect \cyansquare), 0.0438 (\protect \orangediamond), and 0.0985 (\protect \redinvertedtriangle), knowing that $\Gp_c=0.04$. (b,c) Temporal evolution of the size $\ell_b$ of the fluidized band in the case of (b)~permanent shear bands ($\Gp<\Gp_c$) and (c)~transient shear bands ($\Gp>\Gp_c$) for the same shear rates as in (a). The red solid lines represent a logarithmic trend in (b) and the power-law scalings with exponents $1$ and $4/9$ in (c).} \label{fig_7}
\end{center}
\end{figure}

We now discuss the long-time evolution of the stress and velocity profile for the five selected values of $\Gp$ on both sides of the critical shear rate $\Gp_c$. As shown in the inset of Fig.~\ref{figureovershootshearbanding}, after reaching a maximum at times $\tit\simeq 2$--4$\tau$, the stress slowly decreases toward a stationary regime at long times ($\tit>10\tau $). In Fig.~\ref{fig_7}(a), we further report the normalized fluidity $f(y)/m^2$ achieved at long time once a stationary regime is reached, as a function of the normalized position $y/L$. For the three lowest shear rates, which verify $\Gp<\Gp_c$, the shear rate is heterogeneous showing a permanent shear band, i.e., the coexistence between a vacuum and a compact solution~\cite{Benzi:2016}. For shear rates much larger than $\Gp_c$, the fluidity is spatially homogeneous [\protect \redinvertedtriangle~in Fig.~\ref{fig_7}(a)] indicating a homogeneous velocity profile. For shear rates closer to the critical shear rate $\Gp_c$ [\protect \yellowdiamond~in Fig.~\ref{fig_7}(a)], there is no shear band, but the calculated fluidity is not equal to $m^2$ throughout the whole system, which is a signature of the competition between a stable state (here, a homogeneous flow) and an unstable state (shear bands).

Finally, in Figs.~\ref{fig_7}(b) and \ref{fig_7}(c), we report the size of the fluidized band $\ell_b$ as a function of time for shear rates respectively below and above $\Gp_c$. For $\Gp \ge \Gp_c$, Fig.~\ref{fig_7}(c) shows that $\ell_b(\tit)$ follows the same scalings as those predicted in Sect.~\ref{Theory_Homogeneous} in the absence of noise, namely $\ell_b(\tit)\sim (\tit-\titone)$ at short times in the diffusive regime [see Eq.~\eqref{16} and Fig.~\ref{figpreliminary0}(c)] and $\ell_b(\tit)\sim \tit^{4/9}$ at long times [see Eq.~\eqref{lfluid}]. On the other hand, for $\Gp \le \Gp_c$, the shear band tends to grow very slowly after its nucleation, almost logarithmically in time, up to the point where it reaches its asymptotic value [see Fig.~\ref{fig_7}(b)]. In summary, the temporal evolution of the system towards its steady state is not affected by the existence of a stochastic noise in the dynamics as long as $\Gp \ge \Gp_c$.

\section{Comparison with literature data}\label{sec:literatureresults}

In this section, we revisit some previously published data on stress overshoots in a wide variety of glassy systems in light of the present general theory. In the companion paper (see Fig.~3 in Ref.~\cite{companion2}), we successfully compared our theory to experiments on Carbopol microgels that show transient shear banding~\cite{Divoux:2010,Divoux:2011,Divoux:2011b,Divoux:2012}. Equations~\eqref{result_stress_n} and \eqref{alphan} were shown to provide an excellent framework to rescale experimental data onto the two predicted behaviors, namely ({i})~the diffusive low-shear scaling for $\Gp\ll 1$: $\Sm-1\sim\Gp^{\beta(n)}$ with $\beta(n)=2n/3$, and ({ii}) the asymptotic high-shear scaling for $\Gp\gg 1$: $\Sm-1 \sim\Gp^{\alpha(n)}$ with $\alpha(n)=4n/(9-n)$.

\subsection{Overshoot scaling}

\begin{figure}
    \centering
    \includegraphics[width=0.9\columnwidth]{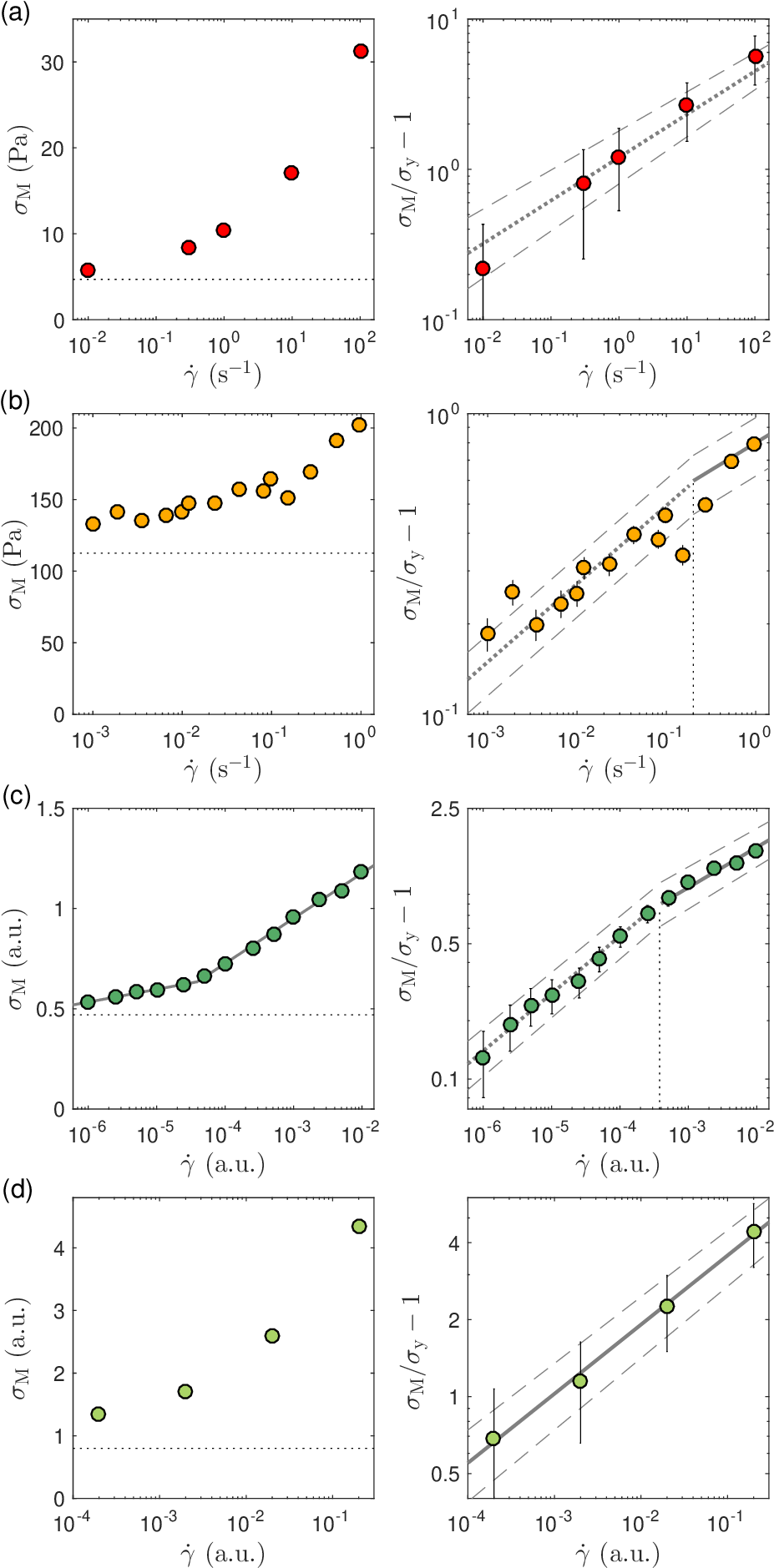}
    \caption{Analysis of stress overshoots reported in the literature for (a)~experiments on a dense assembly of thermosensitive soft core-shell colloids (data from Fig.~3 in Ref.~\cite{Carrier:2009}), (b)~experiments on a commercial hair gel (data from Fig.~4 in Ref.~\cite{Fernandes:2017}), (c)~numerical simulations of a Lennard-Jones glass (data for temperature $T=0.2$ from Fig.~10 in Ref.~\cite{Varnik:2004}), and (d)~numerical simulations of Brownian hard disks (data for volume fraction $\phi=0.81$ from Fig.~9 in Ref.~\cite{Amann:2013}). Left column: stress maximum $\sm$ vs shear rate $\gp$. The dotted lines show the yield stresses. The solid lines in (c) correspond to logarithmic fits $\sm = a+b \log\gp$ at low and high $\Gp$. Right column: rescaled stress maximum as a function of $\gp$. The dotted lines correspond to the scaling predicted in the low-shear (diffusive) regime with exponent $\beta(n)=2n/3$ while the solid lines show the scaling expected in the high-shear regime with exponent $\alpha(n)=4n/(9-n)$, where $n$ is the HB exponent. See Table~\ref{t:parameters} for a list of all various parameters. Error bars correspond to the standard error on the yield stress determination through HB fits. The dashed lines are power laws which exponents reflect the standard error on $n$ as indicated in Table~\ref{t:parameters}. The dotted vertical lines point to the crossover between the diffusive and the asymptotic regimes when present.}
    \label{fig:literature}
\end{figure}

\begin{table*}
\begin{tabular}{c|c|c|c|c|c}
Figure & Ref. & System & $\sy$ (Pa or a.u.) & $n$ & $A$ (Pa.s$^n$ or a.u.) \\
\hline\hline
\ref{fig:literature}(a) &~\cite{Carrier:2009} & core-shell PNIPAM-PS colloids & 4.7$\pm$ 0.7 & 0.43 $\pm$ 0.08 & 2.8 $\pm$ 1.0 \\   
\ref{fig:literature}(b) &~\cite{Fernandes:2017} & hair gel (mostly Carbopol) & 112.6 $\pm$ 1.1  & 0.39 $\pm$ 0.003  & 90.1 $\pm$ 1.4\\   
\ref{fig:literature}(c) &~\cite{Varnik:2004} & 3D 80:20 binary Lennard-Jones mixture & 0.47 $\pm$ 0.01 & 0.445 $\pm$ 0.02  & 2.6 $\pm$ 0.25\\
\ref{fig:literature}(d) &~\cite{Amann:2013} & 2D Brownian hard disks & 0.8 $\pm$ 0.09 & 0.57 $\pm$ 0.02 & 4.8 $\pm$ 0.2\\   
\end{tabular}
\caption{Parameters used in Fig.~\ref{fig:literature}: the yield stress $\sy$, the shear-thinning exponent $n$ and the consistency index $A$ and their standard errors are obtained from HB fits of the data extracted from Fig.~3 in Ref.~\cite{Carrier:2009}, Fig.~1 in Ref.~\cite{Fernandes:2017}, Fig.~2 in Ref.~\cite{Varnik:2004} and Fig.~11 in Ref.~\cite{Amann:2013}.} \label{t:parameters}
\end{table*}
 
We further test the robustness of our modelling of shear start-up based on four additional data sets extracted from the literature, respectively experiments on a dense assembly of soft care-shell colloids~\cite{Carrier:2009} and on a commercial hair gel~\cite{Fernandes:2017} in Fig.~\ref{fig:literature}(a,b) and numerical simulations of a Lennard-Jones glass~\cite{Varnik:2004} and of Brownian hard disks~\cite{Amann:2013} in Fig.~\ref{fig:literature}(c,d). In all cases, the stress maximum increases mildly with the applied shear rate that spans at least three orders of magnitude (left column of Fig.~\ref{fig:literature}). Fitting the steady-state flow curves reported for these various systems by the HB model allows us to extract the corresponding values of the yield stresses $\sy$ and of the HB exponents $n$, as reported in Table~\ref{t:parameters}. When one considers the distance to the yield stress $\sm/\sy-1$, rather than the absolute stress maximum $\sm$ clear power-law behaviors are recovered as a function of the shear rate $\gp$ (right column of Fig.~\ref{fig:literature}). Strikingly, as shown by the lines in Fig.~\ref{fig:literature}, these power laws are in good agreement with the scalings predicted by Eqs.~\eqref{result_stress_n} and \eqref{alphan}, either with exponent $\alpha(n)$ (solid lines) or with exponent $\beta(n)$ (dashed lines), depending on the system and on the shear rate range.

\begin{figure}[b!]
\centering
\includegraphics[width=0.8\columnwidth]{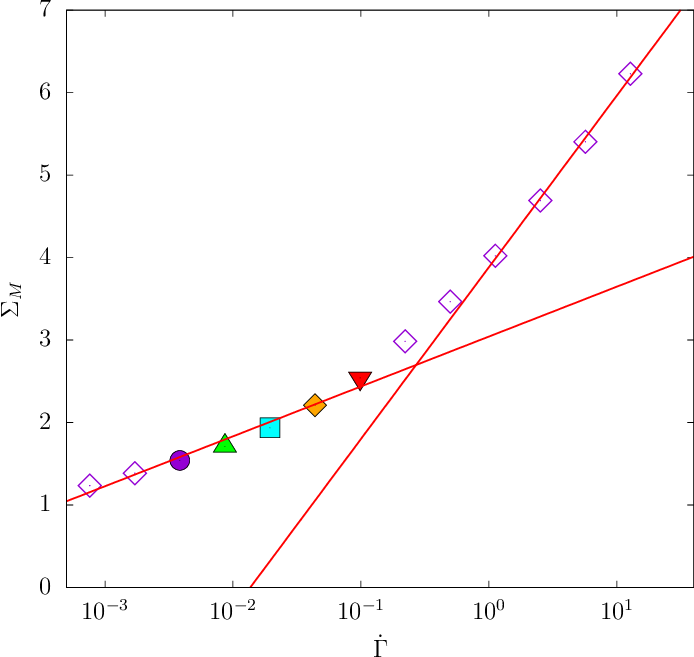}
\caption{Same data $\Sm$ versus $\Gp$ as in Fig.~\ref{figureovershootshearbanding} plotted in semilogarithmic scales. The solid lines are logarithmic fits $\Sm = a+b \log\Gp$ at high and low $\Gp$, respectively. The values of $a$, $b$ are: $a=3.04$, $b=0.26$ (low $\Gp$) and $a=3.88$, $b=0.90$ (high $\Gp$).} 
\label{figure_varnik}
\end{figure}

Comparing the power-law fits for the various data sets, our rescaling shows that the experimental data mostly fall into the diffusive regime [Fig.~\ref{fig:literature}(a,b)], except perhaps for the hair gel at the highest shear rates. On the other hand, the numerical data for the Brownian hard disks are all consistent with the high-shear asymptotic regime [Fig.~\ref{fig:literature}(d)], while those for the Lennard-Jones show a transition from the diffusive to the high-shear scaling for $\dot \gamma \simeq 10^{-4}$ [Fig.~\ref{fig:literature}(c,right)]. The latter result offers a radically different interpretation of the data presented in Ref.~\cite{Varnik:2004}. There, in a case that exhibits steady-state shear banding, the increase of the stress maximum was originally described by two consecutive logarithmic growths $\sm\sim\log\gp$ separated by a critical shear rate that corresponds to the inverse of the structural relaxation time [see solid lines in Fig.~\ref{fig:literature}(c,left)]. Such logarithmic behaviors of $\sm$ with $\gp$ were tentatively related to activated dynamics and shear-induced hopping. In our framework, the same data are rather interpreted as two power-law scalings of $\Sm-1$ vs $\Gp$ in connection with the growth dynamics of the shear band. Such a strong apparent contradiction is resolved in Fig.~\ref{figure_varnik} where the data of Fig.~\ref{figureovershootshearbanding} are shown as $\Sm$ vs $\Gp$ in semilogarithmic scales: two different regimes with apparent logarithmic trends clearly emerge at low and high $\dot \Gamma$, in agreement with the results of Ref.~\cite{Varnik:2004}. Therefore, we may conclude that without any theoretical hint, one can hardly make any strong statement about the overshoot scaling when focusing on $\Sm$ vs $\log\Gp$ or rather on $\Sm-1$ vs $\Gp$. Yet, by accounting for a large amount of experimental and numerical data through power-law scalings, our theory provides an alternative to the interpretation proposed in Ref.~\cite{Varnik:2004}. 

\subsection{Shear band dynamics and fluidization}
\label{sec:SB_longtime}

\begin{figure}[!t]
\centering
\includegraphics[width=0.85\columnwidth]{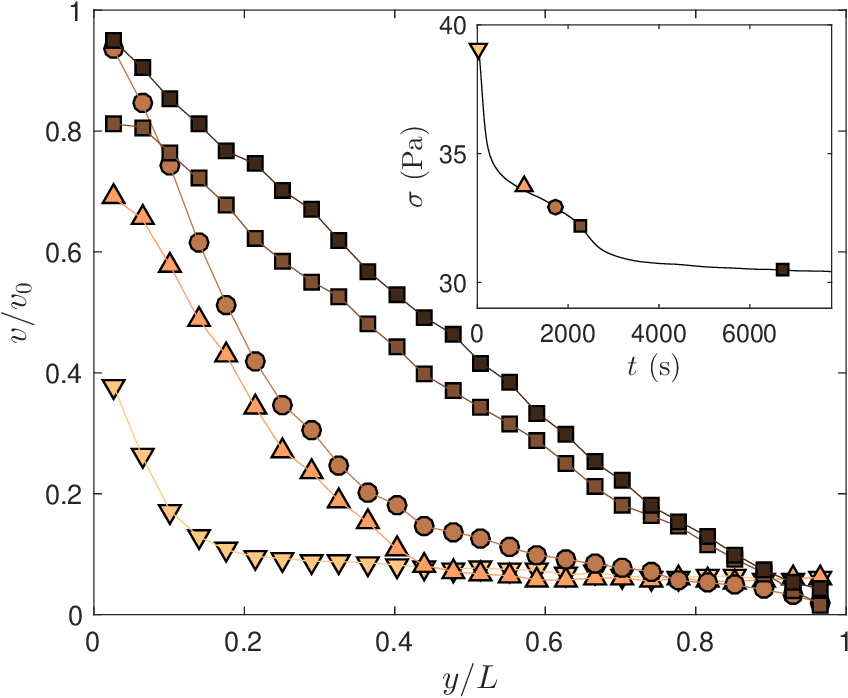}
\caption{Transient shear banding in a 1\%~wt. carbopol microgel. Velocity profiles $v(y)/v_0$ measured by ultrasound velocimetry at different times after a constant shear rate $\gp=0.7$~s$^{-1}$ is applied at $t=0$, which corresponds to a constant velocity $v_0=0.77$~mm\,s$^{-1}$ of the moving wall. Inset: corresponding stress response $\sigma(t)$. The colored symbols show the times at which the velocity profiles in the main graph are recorded, respectively $t/\tau_\text{f}=0.01$, 0.44, 0.75, 0.98, and 2.9, with $\tau_\text{f}=2300$~s. Experiments performed in a concentric-cylinder geometry of gap $L=1.1$~mm covered with sandpaper of roughness 60~$\mu$m (data adapted from Ref.~\cite{Divoux:2010}). }
\label{ffigureAvalanche:1}
\end{figure}

\begin{figure}[t]
\centering
\includegraphics[width=0.85\columnwidth]{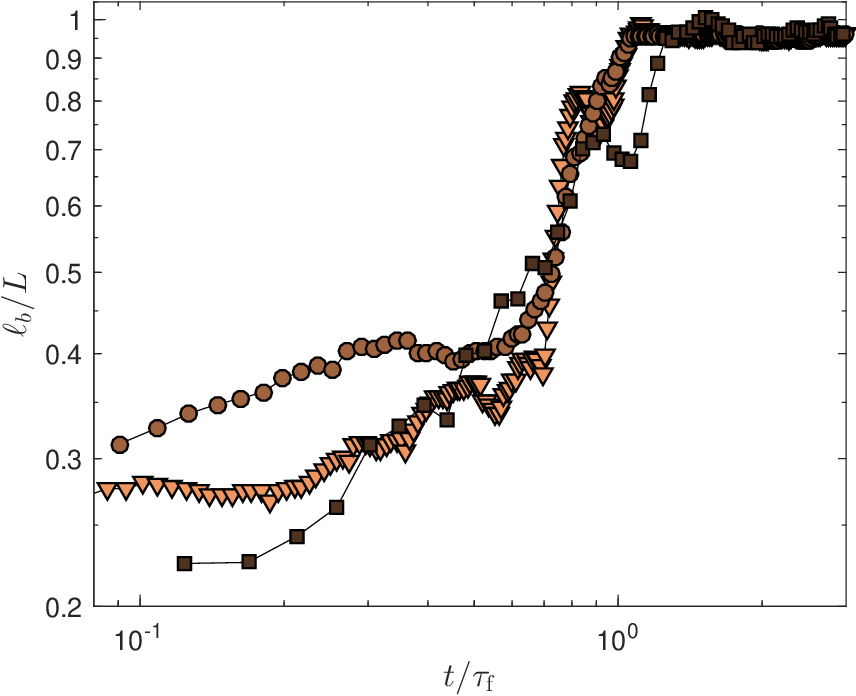}
\caption{Normalized width $\ell_b/L$ of the fluidized band measured as a function of $t/\tau_\text{f}$ in a 1\%~wt. Carbopol microgel during shear start-up at $\gp=0.7$ (\protect \invertedtrianglefigeleven), 1 (\protect \circlefigeleven) and 3~s$^{-1}$ (\protect \squarefigeleven), respectively with $\tau_\text{f}=2300$, 700 and 23~s. Shear is imposed at $t=0$ in a rough concentric-cylinder geometry of gap $L=1.1$~mm covered with sandpaper of roughness 60~$\mu$m (data adapted from Ref.~\cite{Divoux:2010}).}
\label{ffigureAvalanche:2}
\end{figure}

The theoretical scenario described so far not only predicts scaling laws for the stress maximum $\Sm$ at the overshoot but also for the temporal growth of the band $\ell_b(\tit)$. The whole approach hinges on the existence of the cooperative scale $\xi$ that drives the diffusive term $\xi^2\Delta f$ in Eq.~\eqref{6}. In Figs.~\ref{ffigureAvalanche:1} and~\ref{ffigureAvalanche:2}, however, we report some experimental evidence pointing to a more complex scenario than the one solely described in terms of the characteristic scale $\xi$. This will prompt us to generalize our model in the next section in order to account for these experimental observations.

First, Fig.~\ref{ffigureAvalanche:1} shows the experimental velocity profiles recorded in a Carbopol microgel after the stress overshoot, while approaching the steady fluidized state. The fluidized region slowly grows in time, while coexisting with a solid like region where the local shear rate is essentially zero. At some point, just before a small yet detectable drop in the stress response (see inset of Fig.~\ref{ffigureAvalanche:1} for $t\lesssim 2000$~s), we observe a spatial coexistence of the fluidized band with a region with a non-zero shear rate (see brown circles in Fig.~\ref{ffigureAvalanche:1}). Such spatial coexistence is somehow possible only if the fluidized band can propagate its effect on a scale of the order of the system size.

Second, Fig.~\ref{ffigureAvalanche:2} reports the temporal evolution of the size $\ell_b$ of the fluidized band during shear start-up in a Carbopol microgel for various shear rates, including the one used in Fig.~\ref{ffigureAvalanche:1}. Just prior to full fluidization at time $\tau_\text{f}$ and in strong correlation with the stress drop revealed in the inset of Fig.~\ref{ffigureAvalanche:1}, $\ell_b$ is observed to abruptly increase. While the shear band growth displays indications of power-law scaling early after the stress overshoot, such a sudden acceleration is not compatible with the scaling laws of Eqs.~\eqref{16} and \eqref{18}. Note that the experimental fluidization time $\tau_f$ is dimensional, whereas the corresponding theoretical fluidization time $\tilde{\Tf}$ is dimensionless.

The experimental results in Figs.~\ref{ffigureAvalanche:1} and \ref{ffigureAvalanche:2} hint at the presence of ``long-range'' correlations in the system, whose manifestation is an abrupt ``avalanche-like'' change in $\ell_b$ at the end of the fluidization process. At this stage, there is no way for our model for transient shear banding described in Sect.~\ref{Theory_Homogeneous} to reproduce these features. To better emphasize this point, let us consider that shear is started at $t=0$ with the initial condition $f(y,0)=f(0)\ll 1$ \and that the fluidity remains homogeneous, so that the diffusive term in Eq.~\eqref{6} is always zero. Further neglecting the $f^{5/2}$ term and integrating Eq.~\eqref{6} leads to
\begin{equation}
\label{g1}
f(\tit)= \frac{f(0)}{1-f(0)\int_0^{\tit} m(s)\,\dd s}\,.
\end{equation}
Therefore, $f(\tit)/m^2$ becomes of order $1$ on a time scale $(m f(0))^{-1}$, which is extremely large time for a small value of $f(0)$. In other words, the current version of our model misses the possibility for the system to undergo avalanche-like events as shown from the experimental data in Figs.~\ref{ffigureAvalanche:1} and~\ref{ffigureAvalanche:2}. This sets a compelling case for generalizing our model. In the following, we shall add the possibility of avalanches to the fluidization scenario described in the previous sections and explore their consequences on the stress overshoot and on the shear band dynamics.

\section{Model Generalization}\label{sec:generalization}

\subsection{Non-local fluidity model with noise and avalanches}

In the model of Sec.~\ref{Theory_shearbanding}, the introduction of noise was rooted in the presence of plastic rearrangements [see Eqs.~\eqref{phi1}--\eqref{eq:renormalization}]. However, it appears too naive to consider the level of noise as independent of the fluidity. Instead, one expects that once a fluidity fluctuation occurs, this fluctuation could enhance the stochastic perturbations with a positive feedback, which would eventually produce an avalanche-like behavior. We hasten to remark that there are many different ways to modify the noise term in order to account for such a positive feedback. Here, rather than discussing mathematical aspects in detail, we wish to provide simple arguments on how to include avalanche-like behaviour in the dynamical equations.

For simplicity, we shall assume a spatially homogeneous configuration $f(y,\tit)=\phi^2(\tit)$. In the framework of Eq.~\eqref{eq:renormalization}, we consider the following {\it multiplicative} noise term:
\begin{equation}\label{eq:noisephi}
\sqrt{\epsilon_0 + \tilde \epsilon_1 \phi^2(\tit)}\, w(\tit) \,,
\end{equation} 
where $w(\tit)$ is a white noise $\delta$-correlated in time and $\tilde \epsilon_1$ is a positive parameter.

Using the tools of stochastic differential equations, one can show that the effect of such multiplicative noise is to introduce, on average, an instability in the fluidity $f=\phi^2$. More precisely, the noise term given by Eq. (\ref{eq:noisephi}) leads to an additional contribution to the r.h.s.~of Eq.~\eqref{33} that is proportional to $\tilde \epsilon_1 f$, and therefore to an instability, whose growth rate is independent of the initial condition $f(0)$. This result may also be obtained through simple physical considerations by introducing pair-wise 
fluidity interactions in Eq.~\eqref{33}: 
\begin{equation}\label{eq:g2}
m f^2(y)  \rightarrow m f^2(y) + mf(y) \int G(y,y')f(y') \dd y'\,,
\end{equation}
where $G(y,y')$ can be thought of as a {\it random} function taking care of the non-trivial long-range interactions in the system. Next, in the mean-field approximation, we assume that
$m f(y) \int G(y,y')f(y') \dd y'  \sim m f(y) \langle \int G(y,y')f(y') \dd y' \rangle$, where  $\langle\dots\rangle$ stands for the average over all possible random realisations. 
We further estimate the average 
by requiring that it is homogeneous in $y$ and satisfies the scaling transformation given by Eq.~\eqref{10}. This leads to
\begin{equation}\label{eq:g3}
m f(y) \langle \int G(y,y')f(y') \dd y' \rangle 
 \sim \epsilon_1 m^3 f\,,
\end{equation}
where $\epsilon_1$ is a positive parameter that accounts for the averaging procedure. Note that  starting from Eq.~\eqref{eq:noisephi} and using It{\^o} calculus \cite{Schuss:1980}, we would obtain, on average, the exact same result upon identifying $\tilde \epsilon_1 = \epsilon_1 m^3$.

To summarize, whether one thinks in terms of self-sustained noise or in terms of long-range pair-wise fluidity interactions, a simple way to account for avalanche-like behavior is to modify Eq.~\eqref{33} into  
\begin{eqnarray}
\label{m2new}
\frac{\partial f}{\partial \tit} &=&  f  \left[ \xi^ 2  \Delta  f + m f -  f ^ {3/2}  +\epsilon_0/4 \right] \\ 
\nonumber
&+&\epsilon_1 m^3 f\ + \sqrt{\epsilon_0 f} w(y,\tit) \, .
\end{eqnarray}
The parameter $\epsilon_1$ is related to a new time scale, which corresponds to the time needed for long-range interactions to modify the dynamics. In particular, Eq.~\eqref{m2new} shows that $f/m^2$ becomes of order $1$ on a time scale $ (\epsilon_1 m^3)^{-1}$, which is now {\it independent} of the initial condition $f(0)$, contrary to our previous approach [see Eq.~\eqref{g1}]. 
In the following, we mostly focus on the effects of long-range interactions in the case of {\it transient} shear banding, i.e. by solving Eq. (\ref{m2new}) numerically in the case $\epsilon_0=0$. Equation~\eqref{m2new} then reduces to
\begin{equation}
\label{m1new}
\frac{\partial f}{\partial \tit} =  f  \left[ \xi^ 2  \Delta  f + m f -  f ^ {3/2} \right ]+\epsilon_1 m^3 f\,,
\end{equation}
where the case $\epsilon_1=0$ simply corresponds to the model for transient shear banding studied in Sect.~\ref{Theory_Homogeneous} [Eq.~\ref{6}]. The full equation \eqref{m2new} with $\epsilon_0\neq 0$ and $\epsilon_1\neq 0$ will be briefly investigated in Sect.~\ref{sec:fullequation}. As in the previous sections, the main dynamical equation for the fluidity is supplemented with Eqs.~\eqref{4} and \eqref{7}.

\subsection{Avalanche-like dynamics of the fluidized band}

\subsubsection{Working definition of $\ell_b(t)$}

In Sects.~\ref{Theory_Homogeneous} and \ref{Theory_shearbanding}, the size $\ell_b(\tit)$ of the fluidized band could be easily determined at all times as $\ell_b(\tit) = \langle f\rangle(\tit)L/m^2(\tit)$ since the fluidity remains vanishingly small in the solid region. Indeed, in the absence of long-range interactions, i.e., for $\epsilon_1=0$, shear bands, either permanent or transient, imply a ``phase separation" in terms of fluidity between a ``fluid phase" where $f \sim m^2$, and a ``solid phase" where $f \ll m^2$, that are always separated by a sharp interface. 
However, in the presence of long-range interactions, i.e., when $\epsilon_1>0$, the fluidity outside the fluidized band increases exponentially at some point, which reduces the difference between the local shear rates in the two ``phases'', up to the point when a clear separation into two different ``phases'' cannot be recognized any more. To determine such a point and devise an operative definition of $\ell_b(\tit)$, we consider the ratio
\begin{equation}\label{eq:Rgamma}
R_{\gamma}(\tit)=\frac{\Gp_{\rm fluid}(\tit)}{\Gp_{\rm solid}(\tit)}=\frac{m^2(\tit)}{f(L,\tit)}\,, 
\end{equation} 
where $\Gp_{\rm fluid}(\tit)\equiv f(0,\tit) \Sigma(\tit)=m^2(\tit)\Sigma(\tit)$ denotes the local shear rate in the shear band and $\Gp_{\rm solid}\equiv f(y=L,\tit)\Sigma(\tit)$ the local shear rate in the solid phase. Such expressions for the local shear rates in terms of the fluidity are  essentially valid after the overshoot (see Fig.~\ref{figpreliminary1}). $R_{\gamma}(\tit)$ characterizes the spatial phase separation at a given normalized time $\tit$. In particular, we may consider the system as composed of two different phases as long as $R_{\gamma}$ is larger than some value $R^{\star}$. For $R_{\gamma} \le R^{\star}$, the system should be considered as an heterogeneous flow relaxing to its mechanical equilibrium {\it within} the fluidized phases.

The above criterion on $R_{\gamma}$ is equivalent to stating that a clear phase separation occurs as long as $f(L,\tit)\le f^{\star} m^2(\tit)$ where $f^{\star}= 1/R^{\star}$. A simple way to estimate $f^{\star}$ is to consider the r.h.s.~of Eq.~\eqref{m1new}: outside the shear band, the fluidity is homogeneous and the local growth rate of $f$ is given by $2m f+\epsilon_1 m^3 -2.5\,f^{3/2}$. The maximum value of the growth rate is attained for $f=64/225\, m^2 + O(\epsilon_1 m^2) \sim 0.3 m^2$. Therefore, we may choose $f^{\star} = 0.3$ and compute $\ell_b$ from the fluidity profile $f(y,\tit)$ by solving numerically $f(\ell_b(\tit),\tit)=f^{\star} m^2(\tit)$ at each time step of our computations.

\begin{figure}[t]
\centering
\includegraphics[width=1.0\columnwidth]{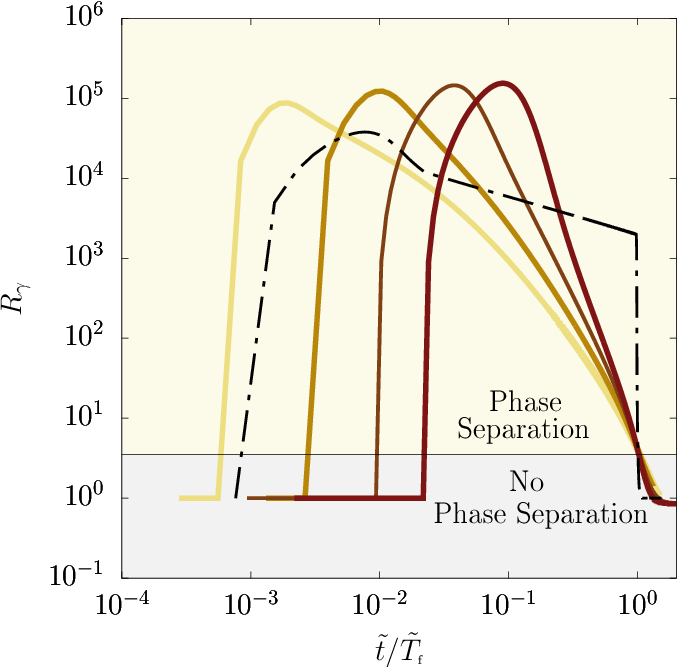}
\caption{Temporal evolution of the ratio $R_{\gamma}$ of the local shear rates as defined by Eq.~\eqref{eq:Rgamma} and computed in the presence of avalanche-like dynamics for $\Gp=0.03$, 0.1, 0.3, and 0.7 from left to right (lighter to darker colors). Numerical results obtained by solving Eq.~\eqref{m1new} with $\epsilon_1=0.1$, $\tau=1$, and $\xi=0.04$. The dashed-dotted line corresponds to $R_{\gamma}(\tit)$ computed for $\epsilon_1=0$ and $\Gp=0.3$. Above the horizontal line at $R_{\gamma}=R^{\star}=1/0.3$, the system separates into a solid phase and a fluid phase. $R_{\gamma}=1$ corresponds to a linear velocity profile. All curves are plotted as a function of the normalized time $\tit/\tilde{\Tf}$ where $\tilde{\Tf}$ is the fluidization strain.}
\label{ffigureAvalanche:7}
\end{figure}

Figure~\ref{ffigureAvalanche:7} shows the temporal evolution of $R_{\gamma}$ obtained by solving Eq.~\eqref{m1new} with $\epsilon_1=0.1$ for  $\Gp = 0.03, 0.1, 0.3, 0.7$ (continuous lines) and compared to the case of transient shear banding without avalanches, i.e. $\epsilon_1=0$, for $\Gp=0.3$ (dash-dotted line). 
In the latter case, once the shear band has nucleated, $R_{\gamma}$ remains always very large, above $10^3$, up to an abrupt jump to $R_{\gamma}=1$ at $\tit=\tilde{\Tf}$ upon fluidization. In the presence of avalanches, however,
$R_{\gamma}$ decreases continuously with a rate that gets faster and faster before reaching $1/f^{\star}$ at the fluidization time.
Indeed, for $\epsilon_1>0$, the model includes two competing mechanisms: the growth of the shear band and the exponential growth of the fluidity outside the shear band. As long as the fluidity outside the shear band remains small, the shear band instability is the driving relaxation mechanism. However, once the fluidity outside the shear band becomes non-negligible, i.e. of the order of $f^{\star} m^2$ with $f^{\star} =0.3$, the avalanche-like event becomes  relevant and the system reaches its final equilibrium in a very short time. This implies that most of the time needed for fluidization is controlled by the time scale of the shear band growth, whereas the final stage occurs, in comparison, on a rather short time. 

\begin{figure}[t]
\centering
\includegraphics[width=1.0\columnwidth]{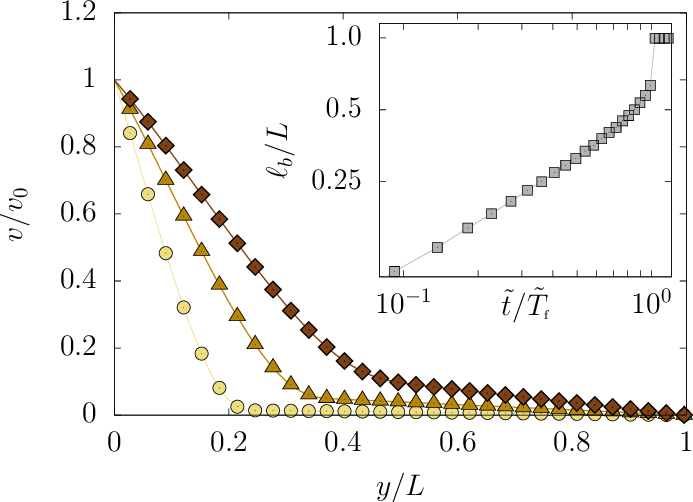}
\caption{Velocity profiles computed at $\tit/\tilde{\Tf}=0.26$ (\protect \circlelightgoldenrod), $0.52$ (\protect \triangledarkgoldenrod) and $0.79$ (\protect \diamondsiennafour) 
in the case of a transient shear band. $\tilde{\Tf}=380$ is the fluidization strain. Profiles are obtained by solving Eq.~\eqref{m1new} for $\epsilon_1=0.1$, $\tau = 1$, $\xi=0.04$ and $\Gp = 0.1$. Inset: $\ell_b$ vs $\tit/\tilde{\Tf}$. \label{ffigureAvalanche:3}}
\end{figure}
Figure~\ref{ffigureAvalanche:3} displays the velocity profiles at three different times in the presence of avalanche-like events for $\epsilon_1=0.1$ and  $\Gp = 0.03$. Prior to complete fluidization, the velocity profile develops a non-negligible shear rate outside the shear band, although we can still distinguish between two regions of different fluidity (see velocity profile for $\tit/\tilde{\Tf}=0.79$ in brown symbols). This behavior is qualitatively very similar to the one reported in experiments (compare with experimental velocity profile for $t/\tau_\text{f}=0.75$ in Fig.~\ref{ffigureAvalanche:1}). Moreover, as seen in the inset of Fig.~\ref{ffigureAvalanche:3}, the size of the shear band $\ell_b(\tit)$ abruptly jumps to $\ell_b=L$ at $\tit =\tilde{\Tf}$. Because of our definition of $\ell_b$, such a jump is expected and corresponds to the point where the fluidity $f(L,\tit)$ reaches $f^\star m^2$.

In order to fully justify our generalized model, we still need to check {(i)} whether the fluidization time $\Tf$ shows the same scaling behaviors for $\epsilon_1>0$ and for $\epsilon_1=0$ and {(ii)} whether there is any evidence for two different fluidization mechanisms, beside our theoretical arguments. We proceed to address these questions in the next two subsections.

\begin{figure}[!t]
\centering
\includegraphics[width=1.0\columnwidth]{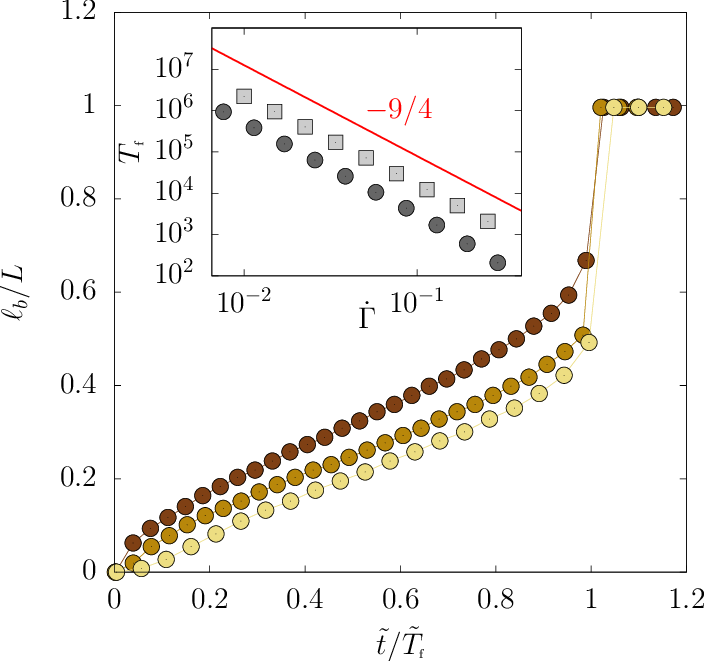}
\caption{Size of the shear band $\ell_b$ as a function of $\tit/\tilde{\Tf}$, where $\tilde{\Tf}$ is the fluidization strain for $\Gp = 0.1$ (\protect \circlelightgoldenrod), $0.3$ (\protect \circledarkgoldenrod), and $0.7$ (\protect \circlesiennafour) versus $\tit/\tilde{\Tf}$. Numerical results obtained by solving Eq.~\eqref{m1new} for $\epsilon_1=0.1$, $\tau=1$, and $\xi=0.04$. Inset: fluidization time $\Tf$ computed for $\epsilon_1=0$ (\protect \graysquare) and $\epsilon_1=0.1$ (\protect \grayscurocircle) versus $\Gp$. The red line shows the power-law scaling $\Tf\sim\Gp^{-9/4}$.}
\label{ffigureAvalanche:4}
\end{figure}

\subsubsection{Fluidization time with avalanches}

Figure~\ref{ffigureAvalanche:4} shows the temporal evolution of $\ell_b$ computed for $\Gp = 0.1$, 0.3, and $0.7$, which all display avalanche-like behavior prior to full fluidization, in qualitative agreement with the experimental data in Fig.~\ref{ffigureAvalanche:2}. Note that our numerical computations cannot account for random-like oscillations as observed experimentally since long-range correlations are parametrized by $\epsilon_1$ using a mean-field approach.

Furthermore, the fluidization times $\Tf$ extracted from our computations are plotted against the imposed shear rate $\Gp$ in the inset of Fig.~\ref{ffigureAvalanche:4} for $\epsilon_1=0$ and $\epsilon_1=0.1$. Obviously, the value of $\Tf$ is smaller for $\epsilon_1>0$ due to avalanches. Yet, in both cases, $\Tf$ is in excellent agreement with the theoretical prediction $\Tf\sim\Gp^{-9/4}$ reported in~\cite{Benzi:2019} and discussed in Sect.~\ref{sec:fluidizationtime}. We conclude that the scaling of $\Tf$ with respect to $\Gp$ is not sensitive to the presence of multiplicative noise and avalanches. This confirms our previous argument based on two widely different competing time scales, which characterize the fast final stage of the fluidization process.

\subsubsection{Signature of avalanches in the stress response}

\begin{figure}[t]
\centering
\includegraphics[width=1\columnwidth]{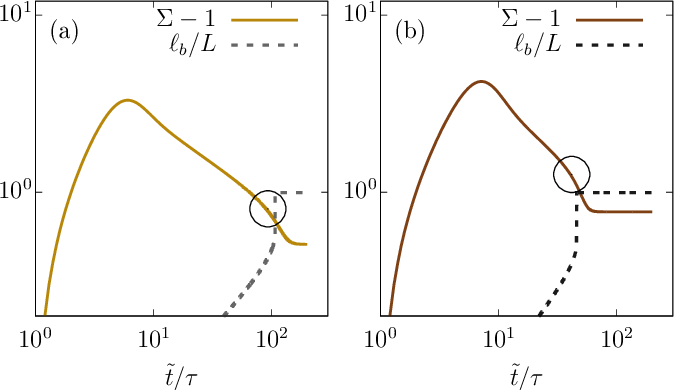}
\caption{Stress response $\Sigma-1$ (solid line) and size of the shear band $\ell_b/L$ (dashed line) versus $\tit/\tau$ for (a) $\Gp=0.3$ and (b) $\Gp =0.7$. The circle highlights a kink after which a fast stress drop occurs. Numerical results obtained by solving Eq.~\eqref{m1new} for $\epsilon_1=0.1$, $\tau=1$, and $\xi=0.04$.}
\label{ffigureAvalanche:6}
\end{figure}

In order to identify a change in the fluidization mechanism associated with avalanches, we focus on the stress response $\Sigma(\tit)$. Figure~\ref{ffigureAvalanche:6} shows $\Sigma-1$ together with $\ell_b/L$ as a function of $\tit/\tau$ for two different applied shear rates. Interestingly, $\Sigma-1$ shows a sharp change of slope around a small kink highlighted with a circle. While $\Sigma$ smoothly decays after the overshoot, it drops much faster and rapidly converges to its steady-state value just after the kink. Such a sudden change and the subsequent quasi-discontinuous transition towards the fully fluidized steady-state stress, for which $\ell_b/L=1$, are also observed in the experimental stress response in the inset of Fig.~\ref{ffigureAvalanche:1}. This hints at a common physical interpretation of the kink in the stress response as the signature of the onset of a new, faster relaxation mechanism in the system, namely the avalanche-like behavior, which leads to a rapid complete fluidization. Therefore, although based on a mean-field approximation, the generalized version of our model introduced in Eq.~\eqref{m1new} captures subtle, yet noticeable and so far unexplained, features of the experiments of Ref.~\cite{Divoux:2010}.

\subsection{Stress overshoot with avalanches: towards brittle-like response}

\subsubsection{Influence of boundary conditions in the presence of avalanches}

In this Section, we explore the role of boundary conditions on our generalized model as defined in Eq.~\eqref{m1new}. We recall that so far we have used the boundary conditions $f(0,\tit)=m^2$ at the shearing wall and $\partial_y f |_{y=L}=0$ at the fixed wall. The boundary condition $f=m^2$ at the moving wall hinges on the physical assumption that, upon increasing the stress, plastic events are triggered by the external driving leading to an increase of the fluidity near the wall. 
One may, however, consider another situation at $y=0$ where $\partial_y f |_{y=0}=0$. Physically, this boundary condition is equivalent to assume that, in the initial stage, the system is so rigid that the driving does not trigger any additional plastic events, even if the stress $\Sigma$ becomes relatively large. At long enough times, if long-range correlations can build up across the system, the growth of the stress may eventually stop. 

\begin{figure}[t]
\centering
\includegraphics[width=1.0\columnwidth]{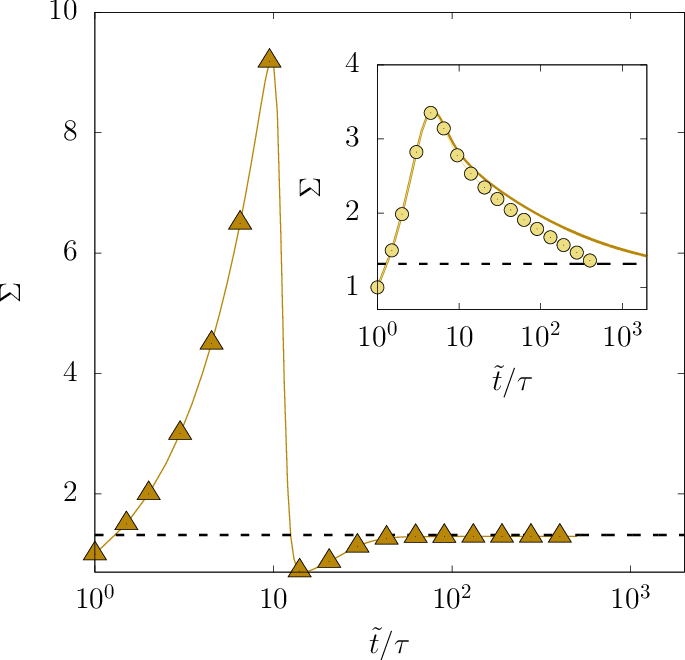}
\caption{Stress response $\Sigma(\tit)$ computed for $\Gp = 0.1$ in the presence of avalanches by solving Eq.~\eqref{m1new} with $\epsilon_1=0.1$, $\tau=1$, $\xi=0.04$, and the boundary condition $\partial_y f |_{y=0}=0$. The horizontal dashed line corresponds to the steady-state HB value $\Sigma_{\rm ss} = 1+\Gp^{1/2}$. Inset: same numerical integration with the boundary condition $f(0,\tit)=m^2$ for $\epsilon_1=0.1$ (\protect \circlelightgoldenrod) and for $\epsilon_1=0$ (solid line).}
\label{ffigureAvalanche:5}
\end{figure}

Figure~\ref{ffigureAvalanche:5} compares the predictions of Eq.~\eqref{m1new} for the two above boundary conditions at the moving wall. As emphasized in the previous section, in the case where the fluidity at the wall is fixed, avalanche-like events become relevant only during the late stages of the shear band evolution, so that we do not expect $\epsilon_1$ to have any significant effect on the scaling of $\Sm$ with $\Gp$. The inset in Fig.~\ref{ffigureAvalanche:5} shows that, when $f(0,\tit)=m^2$, the overshoot is indeed reached for the same value $\Sm$ and at the same time $\titm$ for both $\epsilon_1=0$ and $\epsilon_1=0.1$. We have checked numerically that this result holds for all $\Gp$ and that the scaling of $\Sm$ is still given by Eq.\eqref{general_result} whatever $\epsilon_1$ for this choice of boundary condition.

Remarkably, however, when solving Eq.~\eqref{m1new} under the boundary condition $\partial_y f |_{y=0}=0$ (main panel in Fig.~\ref{ffigureAvalanche:5}), the stress maximum $\Sm$ is much larger than when $f(0,\tit)=m^2$. Moreover, in the former case, the maximum is immediately followed by a very sharp stress drop. This occurs because, for $\partial_y f |_{y=0}=0$, there cannot be any nucleation of a shear band at $y=0$ so that the stress $\Sigma$ grows to rather large values until the whole system is fluidized due to the exponential growth of the initial state, i.e., to the term proportional to $\epsilon_1$ in Eq.~\eqref{m1new}. Just after the overshoot, the abrupt decrease in the stress corresponds to a sharp transition to a fluidized state. Such a behavior is characteristic of a ``brittle'' yielding transition and qualitatively close to the one discussed recently in~\cite{Ozawa:2018,Singh:2020,Barlow:2020}.

Note that in principle one could also explore the effects of the boundary condition $\partial_y f |_{y=0}=0$ on the model discussed in Sect.~\ref{Theory_Homogeneous}, i.e., with $\epsilon_1=0$. However, as discussed in Sect.~\ref{sec:SB_longtime}, with a homogeneous initial condition $f(y,0)=f(0)\ll 1$, the time to reach a fluidity that is large enough for the stress overshoot to occur is proportional to $1/f(0)$ [see Eq.~\eqref{g1}]. Imposing $\partial_y f |_{y=0}=0$ in this case would thus lead to a rather non-physical situation characterized by extremely large values of the stress overshoot occurring at extremely large values of the strain (i.e., of the time $\tit$). Therefore, in the following, we shall not investigate this possibility.

\subsubsection{Scaling of the stress overshoot}
\label{sec:overshoot_brittle}

\begin{figure}[t!]
\begin{center}
\includegraphics[width=1.0\columnwidth]{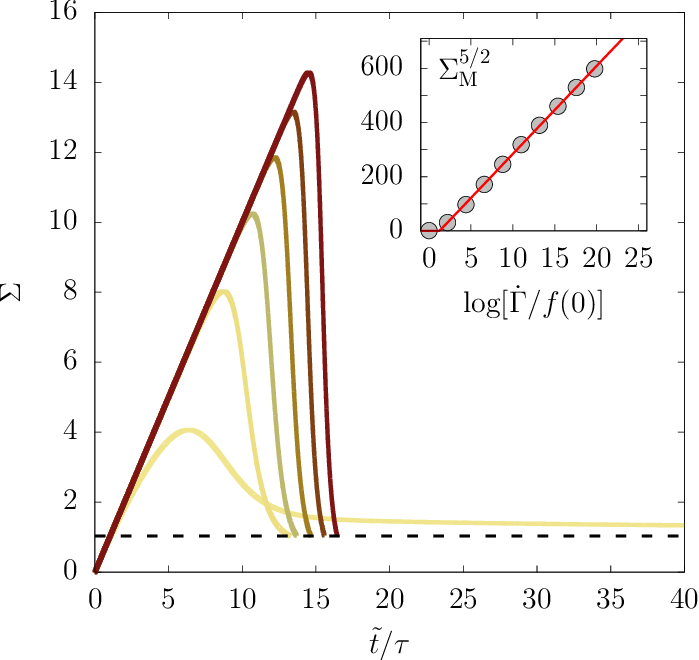}
\end{center}
\caption{Stress response $\Sigma(\tit)$ computed for $\Gp=10^{-3}$ and $f(0)=10^{-4}$, $10^{-6}$, $ 10^{-8}$, $10^{-10}$, $  10^{-12}$, and $10^{-14}$ from left to right (lighter to darker colors). The horizontal dashed line is the steady-state HB value, i.e., $\Sigma_\text{ss}= 1+\Gp^{1/2}$. Numerical results obtained by solving Eq.~\eqref{m1new} for $\epsilon_1=0.1$, $\tau=1$, $\xi=0.04$, and the boundary condition $\partial_y f |_{y=0}=0$. The computation is stopped when the HB value is reached. Inset:  $\Sm^{5/2}$ as a function of $ \log[\dot{\Gamma}/f(0)]$. The red solid line shows the linear behavior predicted by Eq.~\eqref{g4}.}
\label{figgeneral2}
\end{figure}
Let us now investigate the scaling properties of the stress overshoot when $\epsilon_1>0$ and $\partial_y f |_{y=0}=\partial_y f |_{y=L}=0$. Starting with a  homogeneous state $f(y,0) = f(0)\ll 1$, the system cannot develop any spatial heterogeneity in $f(y,\tit)$. Therefore, we can rewrite Eqs.~\eqref{m1new} and \eqref{7} respectively as:
\begin{eqnarray}\label{simple_1}
\frac{\dd f}{\dd\tit}&=& m f^2 + \epsilon_1 m^3 f - f^{5/2}\,,\\
\label{simple_2}
\frac{\dd\Sigma}{\dd\tit} &=& \frac{1}{\tau}  \left( 1- \frac{f\Sigma}{\Gp} \right)\,.
\end{eqnarray}
Assuming $f\ll 1$, we may neglect the terms $mf^2$ and $f^{5/2}$ in Eq.~\eqref{simple_1}.  The expression of $m(\tit)$ given by Eq.~\eqref{12} then leads to
\begin{equation}
    \label{ftf0}
    f(\tit) \sim f(0) \exp\left[ \frac{2\epsilon_1\tau}{5} \left(\frac{\tit-\titone}{\tau} \right)^{5/2} \right]\,.
\end{equation}
The balance equation $f(\titm) \Sm = \Gp$ in Eq.~\eqref{13} further yields
\begin{eqnarray}
\left( \frac{\titm-\titone}{\tau}\right) &\sim&  \left[   \frac{ 5}{2 \tau \epsilon_1} \log \left(\frac{\Gp}{f(0)\Sm}\right) \right]^{2/5} \nonumber \\
&\sim& \left[   \frac{ 5}{2 \tau \epsilon_1} \log \left(\frac{\Gp}{f(0)}\right) \right]^{2/5}\,.
\label{tM} 
\end{eqnarray}
where the last expression neglects a logarithmic correction in $\Sm$. Finally, Eq.(\ref{tM}) implies
\begin{equation}
\label{g4}
\Sm \sim  \left[   \frac{ 5}{2 \tau \epsilon_1} \log \left(\frac{\Gp}{f(0)}\right) \right]^{2/5}\,.
\end{equation}
The above expression allows us to compute the rate of the stress drop $-\dd\Sigma^\star/\dd \tit = \sup_{\tit>\titm} [-\dd\Sigma/\dd \tit \,] $ after the overshoot. Since the stress drop is of the order of $\Sm$ and occurs on a time scale $m^{-3}$, using Eq.~\eqref{12}, \eqref{tM}, and \eqref{g4} leads to 
\begin{equation}\label{ratedrop}
-\frac{\dd\Sigma^\star}{\dd \tit} \sim \Sm m^{3} \sim  \log \left( \frac{\Gp}{f(0)}  \right)\,.
\end{equation}

Figure~\ref{figgeneral2} presents the stress responses $\Sigma$ as a function of the normalized strain $\tit/\tau$ for different values of $f(0) \in [10^{-14},10^{-4} ] $ and for a given shear rate $\Gp = 10^{-3}$. A smaller initial fluidity leads to a larger stress overshoot that occurs at larger strains and with larger decay rate. Such an enhancement of the ``brittleness'' can be rationalized qualitatively if one interprets $f(0)$ as linked to the system rigidity at $\tit=0$. Indeed, the smaller $f(0)$, the smaller the rate of plastic events at $\tit=0$. For a system where plastic events are triggered by a (real or effective) temperature $T$, one expects $f(0) \sim \exp(-E/T)$, where $E$ is the typical energy barrier for plastic events to occur. In this case, $\log(1/f(0)) \sim 1/T$ so that decreasing $f(0)$ corresponds to lowering the temperature. For a system showing physical aging, one rather expects $f(0) \sim t_w ^{-a}$, where $t_w$ is the waiting time and $0<a<1$ \cite{Joshi:2018}. In this case, $\log(1/f(0)) \sim a \log(t_w)$ and the limit $f(0) \rightarrow 0$ corresponds to infinite waiting times. In both cases, a decrease in $f(0)$ amounts to an increase in the initial ``stiffness'' or ``rigidity'' of the system.

Thus, Fig.~\ref{figgeneral2} is consistent with the fact that both the stress overshoot $\Sm$ and its sharpness $-\dd\Sigma^\star/\dd \tit$ increase with increasing initial rigidity for a given $\Gp$. While the stress drop is very sharp and typical of a ``brittle'' transition for small $f(0)$, the stress shows a rather smooth decay after the overshoot for the largest initial fluidity $f(0)=10^{-4}$, which indicative of a ``ductile'' transition. More quantitatively, the inset in Fig.~\ref{figgeneral2} shows that Eq.~\eqref{g4} provides an excellent prediction for the power-law behavior of $\Sm$ as a function of $\log[\Gp/f(0)]$.

\begin{figure}[t!]
\begin{center}
\includegraphics[width=1.0\columnwidth]{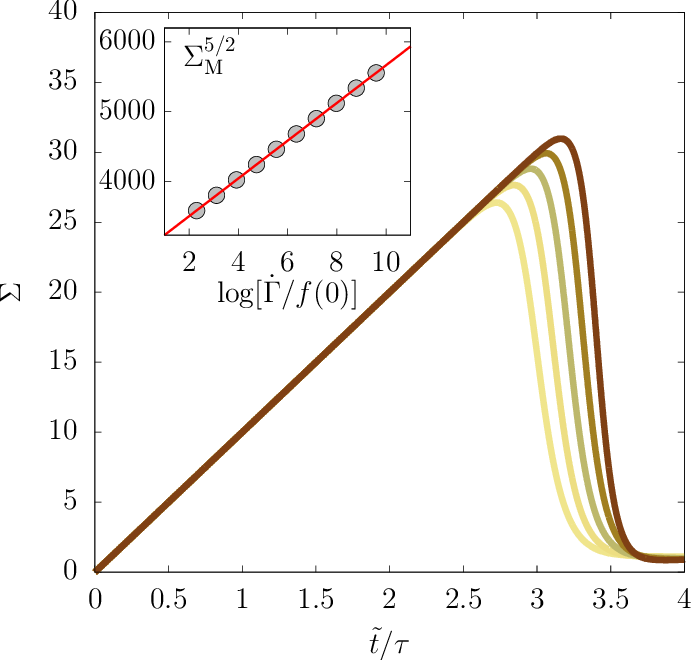}
\end{center}
\caption{Stress response $\Sigma(\tit)$ computed for $f(0)=10^{-4}$ and $\Gp= 0.001$, $0.005$, $0.025$, $0.13$, and $0.66$ from left to right (lighter to darker colors). Numerical results obtained by solving Eq.~\eqref{m1new} for $\epsilon_1=0.1$, $\tau=1$, $\xi=0.04$, and the boundary condition $\partial_y f |_{y=0}=0$. Inset: $\Sm^{5/2} $  as a function of $ \log(\dot{\Gamma}/f(0))$. The red solid line shows the linear behavior predicted by Eq.~\eqref{g4}.}
\label{figgeneral3}
\end{figure}

To complete our analysis, we report in Fig.~\ref{figgeneral3} the evolution of $\Sigma$ as a function of strain for different values of $ \Gp$ and for a given initial fluidity $f(0)=10^{-4}$. An excellent agreement with Eq.~\eqref{g4} is observed over the whole range of shear rates under study for the scaling of $\Sm^{5/2}$ as a function of $ \log[\dot{\Gp}/f(0)]$ (see inset in Fig.~\ref{figgeneral3}). We emphasize that all the results discussed in the previous sections do not depend on the choice of $f(0)$ as long as $f(0)>0$. Here, the situation is quite different: when shear band nucleation due to external forcing cannot occur, the dependence of the stress maximum $\Sm$ on both the initial condition $f(0)$ and the shear rate $\Gp$ strongly differs from the one predicted by Eq.~\eqref{general_result_n}. 

\begin{figure}[t]
\centering
\includegraphics[width=0.9\columnwidth]{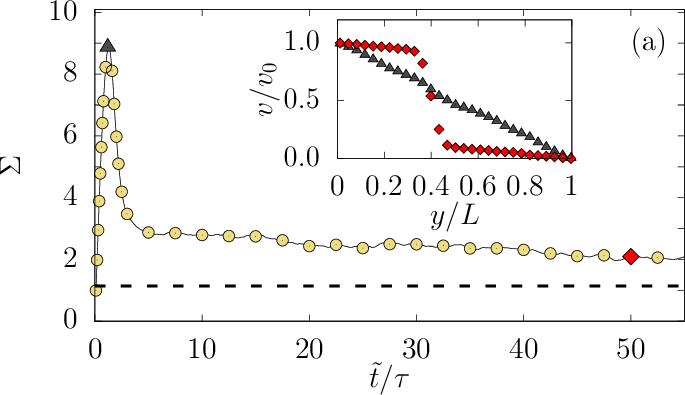}
\includegraphics[width=0.9\columnwidth]{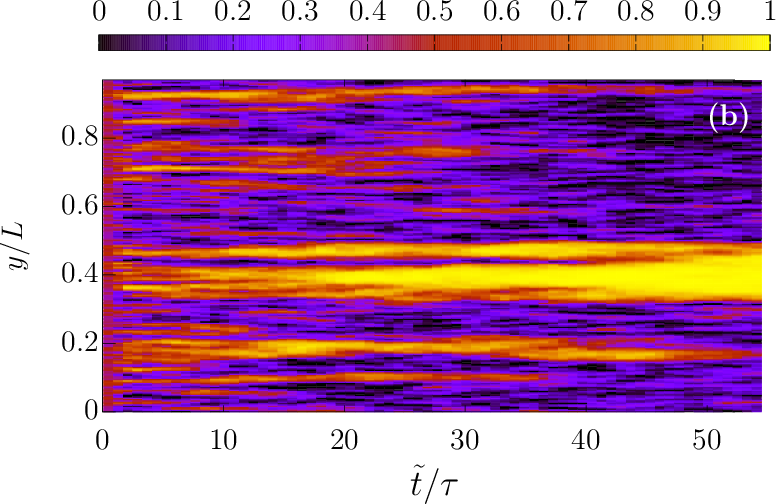}
\caption{(a) Stress response $\Sigma(\tit)$ computed for $\Gp=0.02$ by solving Eq.~\eqref{m2new} with $\epsilon_0 = 10^{-6}$, $\epsilon_1=0.1$, $\tau=1$, $\xi=0.04$, and the boundary condition $\partial_y f |_{y=0}=0$. Permanent shear bands exist for $\Gp \le \Gp_c \simeq 0.1$ (see section~\ref{Theory_shearbanding}). The horizontal dashed line is the steady-state HB value, i.e., $\Sigma_\text{ss}= 1+\Gp^{1/2}$. Inset: velocity profiles at $\tit/\tau=1$ (\protect \graytriangle) and $\tit/\tau=50$ (\protect \reddiamond). (b)~Spatiotemporal diagram of $f(y,\tit)/m^2$ rescaled between 0 and 1 (see color bar). }
\label{ffigureAvalanche:8}
\end{figure}

\subsubsection{Combined effect of avalanches and boundary conditions in the case of steady-state shear banding}
\label{sec:fullequation}

Let us close this section by investigating the effect of both $\epsilon_1$ and the boundary conditions on the formation of a {\it permanent} shear band, i.e., on the model introduced in Sect.~\ref{Theory_shearbanding} where $\epsilon_0>0$. To this aim, Eq.~\eqref{m2new} is supplemented with the boundary conditions $\partial_y f |_{y=0}=\partial_y f |_{y=L}=0$. Because of the space-dependent noise, the effect of cooperativity, i.e., the Laplacian term in the fluidity equation, can no longer be disregarded. Figure~\ref{ffigureAvalanche:8} displays the results of the full model for $\Gp=0.02$, $\epsilon_0 = 10^{-6}$, and $\epsilon_1=0.1$. The stress response $\Sigma(\tit)$ shows a large overshoot followed by a quite sharp stress drop [Fig.~\ref{ffigureAvalanche:8}(a)]. The velocity profiles remain linear up to the stress overshoot, after which the system cannot reach a fully fluidized state. Because $\Gp < \Gp_c$, the system rather develops a steady shear band at $y/L\simeq 0.4$ at long times [inset of Fig.~\ref{ffigureAvalanche:8}(a)]. As shown in the spatiotemporal diagram of Fig.~\ref{ffigureAvalanche:8}(b), the fluidity field $f(y,t)$ normalized by $m^2$ is constituted of a large number of small shear bands (with large fluidity) that spontaneously emerge just after the overshoot and progressively give way to only one large shear band that survives in the middle of the system at long enough times. In this case, our model leads to a clear phase separation {\it long after} the stress overshoot, associated with the formation of a permanent shear band. It is interesting to notice that similar observations were reported in the numerical simulations of Ref.~\cite{Ozawa:2018}.

To summarize, although extremely simplified, the model generalization introduced in Eq.~\eqref{m1new} seems to capture most of the phenomenology reported in the experiments of Figs.~\ref{ffigureAvalanche:1} and \ref{ffigureAvalanche:2}. Moreover, within a single theoretical framework, our model accounts  qualitatively for ``brittle'' and ``ductile'' yielding transitions depending on different choices of the boundary conditions. Remarkably, rather different behaviors, including stress overshoots, transient or steady-state shear banding, avalanche-like fluidization, and brittle-like yielding may thus be explained based on the same theoretical approach.

\section{Discussion and conclusion}\label{sec:discussion_conclusion}

In this manuscript, we have used a continuum model to describe the shear-induced solid-to-liquid transition of YSFs. The present work builds upon a kinetic elastoplastic model, originally derived by Bocquet {\it et al.}~in \cite{Bocquet:2009} to describe the stationary flow of soft glassy materials. Our results constitute a significant improvement over the original model, as our approach accounts for the \textit{transient} response towards steady state, including the presence of shear bands, i.e., the coexistence in space and time of two different phases, namely a solid phase and a fluidized phase. Our fluidity approach, which is based on a dynamical equation coupled to a Maxwell model, includes four control parameters, namely the cooperativity length $\xi$, the Maxwell relaxation time $\tau$, and the amplitudes of two stochastic noises $\epsilon_0$ and $\epsilon_1$, which correspond to short-range and long-range fluidity fluctuations, respectively. 

In practice, starting from the solid state, the solid-to-liquid transition under applied shear rate is associated with a stress overshoot. At the mesoscale, the yielding of the material corresponds to a phase transition that comes in two flavors depending on the boundary conditions at the moving wall and on the noise amplitude.

On the one hand, fixing the fluidity at the moving wall, the short-term rheological response is dictated by the nucleation and growth of a fluidized band in the vicinity of that boundary. This result is robust, independent of the presence of both short-range and long-range noises. In other words, the stress overshoot characteristics are essentially independent of the fate of the shear band, whether stable or unstable. Moreover, the locus of the stress maximum scales as a power-law with the applied shear rate, whose exponent depends on both the HB exponent $n$ and on the short-time fluidization dynamics, which involves either a diffusive growth at small shear rates or a front propagation at large shear rates. In case the shear band is transient, its lifespan decreases for increasing shear rates and scales as $\Tf\sim \gp^{-9/4}$, which exponent is universal and insensitive to both parameters $\epsilon_0$ and $\epsilon_1$. 

On the other hand, setting the fluidity gradient to zero at the moving wall prevents the growth of a shear band at this boundary. In the presence of short-range and long-range noises, this condition results in diffuse plastic events in the bulk, leading to multiple shear bands that compete with one another. The macroscopic signature of this scenario is seen in both the stress maximum that shows a smoother logarithmic dependence with the applied shear rate and in the subsequent stress relaxation, which is characterized by a sharp and almost discontinuous drop that resembles that of a brittle failure. The presence of long-range correlations in the fluidity via the parameter $\epsilon_1$ also results in a sudden, avalanche-like fluidization at long times.  

These scenarios show that ductile and brittle transitions are two natural outcomes of our fluidity model, whose main feature is to allow for the coexistence of two phases. Our model's predictions are in quantitative agreement with a broad collection of numerical and experimental results, encompassing colloidal gels and soft glasses. In that sense, our approach unifies many different features of transient flows in soft glassy materials. Nonetheless, our approach neglects some features that may play an important role during shear start-up flows, including lubrication forces. Such elastohydrodynamic (EHD) interactions \cite{Seth:2008,Seth:2011,Liu:2018c} were recently shown to affect the scaling of the stress maximum reached during the stress overshoot \cite{Khabaz:2021}. Our model is versatile enough that an EHD contribution could be easily included in our approach, which is left for future work.      

Finally, with respect to previous experimental and numerical work on the ductile-to-brittle transition in soft materials \cite{Picallo:2010,Tixier:2010,Ramos:2011,Erk:2012,Arif:2012,Kovacs:2013,Barlow:2020}, a novel feature highlighted by our study is that boundary conditions may control such a transition in soft glassy materials. We have used the simplest boundary conditions that allow  or suppress phase nucleation at the moving wall. Yet, a clear physical connection between boundary conditions and the microscopic processes at the walls is still missing. Indeed, in experiments, such processes may depend on numerous factors such as the wall surface roughness or the interaction between the sample and the wall. Therefore, future work should focus on providing a microscopic justification for the various types of boundary conditions used in the present work.

\section{Appendix}
\label{app:lsa}
In this Appendix, we provide supplemental information on the velocity of the fluidized band and on its time dependence. 
The strategy is the following: {(i)} we first recall some basic features on a relevant reaction-diffusion equation and the velocity of its travelling fronts; {(ii)} we then use these results to capture some interesting information about the scaling of the front velocity $\dd\ell_b/\dd\tit$ as a function of $m$ {and $\tit$}. 
\subsection{Travelling fronts in case of a linear reaction-diffusion equation}\label{app:sub1}
We consider the scalar field $f$ obeying the following linear reaction-diffusion equation
\begin{equation}\label{eq:simpleREF}
\frac{\partial f}{\partial \tit} = D \Delta f + m f \,,
\end{equation}
where $D$ is a diffusion constant and $m$ is a reaction constant. It is convenient to rewrite the equation as 
\begin{equation}
\frac{\partial }{\partial \tit} \left[ \exp \left(-m \tit \right) f \right] =   D \Delta \left[ \exp \left( - m \tit \right) f \right]\,,
\end{equation}
so that $\exp\left(-m \tit \right) f$ is solution of the diffusion equation, i.e., $\exp \left(- m \tit \right) f \sim \exp\left( -\frac{y^2}{4 D \tit}\right)$, hence
\begin{equation}\label{eq:simpleRDE}
\begin{split}
f & \sim \exp \left[ m \tit-\frac{y^2}{4 D \tit} \right] \\
  & = \exp \left[ \frac{1}{\tit}\left( \sqrt{m} \tit+\frac{y}{\sqrt{4D}} \right) \left( \sqrt{m} \tit -\frac{y}{\sqrt{4D}} \right) \right].
\end{split}
\end{equation}
These calculations show that the solution of Eq.~\eqref{eq:simpleREF} exhibits a travelling-wave behaviour with a characteristic velocity
\begin{equation}\label{eq.VELOCITYFRONT}
|v_f| \sim \sqrt{D m}\,.
\end{equation}
{If the reaction constant $m$ depends on time and provided the characteristic time scale of the evolution of $m$ is large compared to the one of $f$, we can use a quasi-stationary approximation and extend the previous result to approximate the time evolution of the front velocity by
\begin{equation}\label{eq.VELOCITYFRONTsimple}
|v_f(\tit )| \sim \sqrt{D m(\tit )}\,.
\end{equation}
}
\subsection{Travelling Fronts in case of the fluidity model}
Let us now consider the following equation for the fluidity
\begin{equation}\label{eq:fluidityWAVE}
\frac{\partial f}{\partial \tit} = f  \left[\xi^2 \Delta f + m f - f^{3/2} \right]\,,
\end{equation}
where $\xi$ is constant and $m$ time-dependent. 
{Let us now assume that $m(\tit)$ is a ``mild'' function of time $\tit$ in comparison to the velocity of the fluidized band $\dd\ell_b/\dd \tit$.} 
This assumption allows us to solve  Eq.~\eqref{eq:fluidityWAVE} with $m$ constant and to determine the scaling of the front velocity $\dd\ell_b/\dd \tit$ as a function of $m$. We assume that at some reference time (say $\tit=0$), the system is characterized by the following initial profile
\begin{equation}
\label{eq:initialcondition}
f(y,0)=\begin{cases} f_{\mbox{\tiny 0}}(y) & \hspace{.2in} 0 \le y \le \ell_b(0) \\
0 & \hspace{.2in} \ell_b(0) \le y \le L\,, \end{cases}
\end{equation}
where $\ell_b(0)=\ell_b(\tit=0)$ is the initial width of the fluidized band and where $f_{\mbox{\tiny 0}}(y)$ is some space-dependent solution which satisfies the stationary equation:
\begin{equation}\label{eq:COMPACTON}
f_{\mbox{\tiny 0}} \left[\xi^2 \Delta f_{\mbox{\tiny 0}} + m f_{\mbox{\tiny 0}} - f_{\mbox{\tiny 0}}^{3/2} \right] = 0\,.
\end{equation}  
We then look for the dynamical equation of a perturbation field $\delta f(y,\tit)$ that is space- and time-dependent, and which is non-zero outside the shear band. To this aim, we use the following profile $f(y,\tit)=f(y,0)+\delta f (y,\tit)$ in Eq.~\eqref{eq:fluidityWAVE}.

Using Eq.~\eqref{eq:COMPACTON}, we find at first order in $\delta f$ that the perturbation field satisfies 
\begin{equation}\label{eq:perturb}
\frac{\partial (\delta f)}{\partial \tit}=\xi^2 f_{\mbox{\tiny 0}} \Delta \delta f + \left(m f_{\mbox{\tiny 0}}-\frac{3}{2} f^{3/2}_{\mbox{\tiny 0}}\right) \delta f\,,
\end{equation}
where $f_{\mbox{\tiny 0}}$ is a  space-dependent function. Hence, both terms $\xi^2f_{\mbox{\tiny 0}}$ and $\left(m f_{\mbox{\tiny 0}} - \frac{3}{2} f^{3/2}_{\mbox{\tiny 0}}\right)$ in Eq.~\eqref{eq:perturb} are non constant in space. 
{This implies that the front velocity is also space-dependent [see Eq.~\eqref{eq.VELOCITYFRONTsimple}].} 
In order to go one step further, we assume that the velocity of the fluidized band  is controlled by the maximum front velocity.  This is a reasonable approximation that we expect not to impact the scaling properties of the front velocity as a function of $m$. Let us denote $f^{(M)}_{\mbox{\tiny 0}}$ the maximum fluidity such that the new reaction-diffusion equation to solve is 
\begin{equation}\label{eq:almostDONE}
\frac{\partial (\delta f)}{\partial \tit} = \xi^2 f^{(M)}_{\mbox{\tiny 0}} \Delta \delta f + \left(m f^{(M)}_{\mbox{\tiny 0}}-\frac{3}{2} (f^{(M)}_{\mbox{\tiny 0}})^{3/2} \right) \delta f\,.
\end{equation}
Based on the results of Eq.~\eqref{eq.VELOCITYFRONTsimple} and assuming that $m$ does not change much at the onset of the instability, we find that the front velocity is given by
\begin{equation}\label{eq:SCALINGFRONT}
\frac{\dd \ell_b}{\dd \tit} \sim \xi \sqrt{ f^{(M)}_{\mbox{\tiny 0}} \left(m f^{(M)}_{\mbox{\tiny 0}}-\frac{3}{2} (f^{(M)}_{\mbox{\tiny 0}})^{3/2} \right)}\,.
\end{equation}
Hence, assuming that $f^{(M)}_{\mbox{\tiny 0}} \sim m^2$, we get the following scaling
\begin{equation}\label{eq:SCALING}
\frac{\dd \ell_b}{\dd \tit} \sim \xi m^{5/2}\,.
\end{equation}
Note that the initial condition [Eq.\eqref{eq:initialcondition}] can also be supplemented with a constant fluidity in space $f_0 \ll 1$, which may be related to the way the sample is prepared. Numerical computations showed that the result reported in Eq.~\eqref{eq:SCALINGFRONT} is robust with respect to changes in the value of $f_0$. In particular, we found that the prediction for the front velocity in Eq.~\eqref{eq:SCALINGFRONT} holds up to logarithmically small corrections.

In the above derivations, we assume the existence of a propagating front, and consequently we relate the front velocity to $m$ by a suitable scaling. However, the solution may not necessarily be that of a propagating front. As discussed in the text, the early time dynamics rather result from a diffusive process allowing the system to adapt to the boundary conditions. If diffusion takes place, the scaling for $\ell_b(\tit)$ is different. Indeed, looking back at Eq.~\eqref{eq:fluidityWAVE} and concentrating only on the balance of 
\begin{equation} 
\frac{\partial f}{\partial \tit} \sim f  \xi^2 \Delta f \,,
\end{equation}
the term $f \xi^2$ appears as a diffusion coefficient. If we assume $f \sim m^{2}$, we get the following prediction for the diffusion-driven growth of the fluidized band
\begin{equation}\label{eq:SCALING_DIFFUSION}
\ell_b(\tit) \sim \xi\, m\, \tit^{1/2}\,.
\end{equation}
{Thus, depending on the physical mechanism responsible for the growth of the fluidized band, we find different behaviours for $\ell_b(\tit)$ namely Eq.~\eqref{eq:SCALING} in the case of front propagation and Eq.~\eqref{eq:SCALING_DIFFUSION} in the case of diffusion.}

\begin{acknowledgments}
This research was supported in part by the National Science Foundation under Grant No. NSF PHY-1748958 through the KITP program on the Physics of Dense Suspensions. This work received funding from the European Research Council (ERC) under the European Union's Horizon 2020 research and innovation
programme (grant agreement No 882340).
\end{acknowledgments}

\bibliographystyle{apsrev4-1}

%

\end{document}